\documentclass[%
floatfix,
showpacs,
nofootinbib,
superscriptaddress
]{revtex4}
\usepackage{amsmath,amssymb,braket,bm,aas_macros}
\usepackage[dvipdfmx]{graphicx}
\usepackage{ulem}
\input{colordvi.tex}
\usepackage{color}
\begin{document}
\title{Impact of anisotropic stress of free-streaming particles
on gravitational waves induced by cosmological density perturbations}

\author{Shohei Saga}
\affiliation{Department of Physics and Astrophysics, Nagoya University,
Aichi 464-8602, Japan}
\email{saga.shohei@nagoya-u.jp}
\author{Kiyotomo Ichiki}
\affiliation{Department of Physics and Astrophysics, Nagoya University,
Aichi 464-8602, Japan}
\affiliation{Kobayashi-Maskawa Institute for the Origin of Particles and the Universe, Nagoya University,
Nagoya 464-8602, Japan}
\author{Naoshi Sugiyama}
\affiliation{Department of Physics and Astrophysics, Nagoya University,
Aichi 464-8602, Japan}
\affiliation{Kobayashi-Maskawa Institute for the Origin of Particles and the Universe, Nagoya University,
Nagoya 464-8602, Japan}
\affiliation{Kavli Institute for the Physics and Mathematics of the Universe (Kavli IPMU), The University of Tokyo,
Chiba 277-8582, Japan}

\begin{abstract}
 Gravitational waves (GWs) are inevitably induced at second-order in cosmological perturbations through non-linear couplings with first order 
 scalar perturbations, whose existence is well established by recent
 cosmological observations. So far, the evolution and the
 spectrum of the secondary induced GWs have been derived by taking into
 account the sources 
 of GWs only from the product of first order scalar perturbations. Here we newly
 investigate the effects of purely second-order anisotropic stresses of
 photons and neutrinos on the evolution of GWs, which have been omitted
 in the literature. We present a full treatment of the 
 Einstein-Boltzmann system to calculate the spectrum of GWs with
 anisotropic stress based on the formalism of the cosmological
 perturbation theory. 
We find that photon anisotropic stress amplifies
 the amplitude of GWs by about $150 \%$ whereas neutrino anisotropic stress suppress that of GWs by about $30 \%$ on small scales $k\gtrsim
 1.0~h{\rm Mpc}^{-1}$ compared to the case without anisotropic
 stress.
 The second order anisotropic stress does not affect GWs with wavenumbers $k\lesssim
 1.0~h{\rm Mpc}^{-1}$.
 The result is in marked contrast with the case at linear order, where
 the effect of anisotropic stress is damping in amplitude of GWs.
 \end{abstract}
\maketitle
\section{Introduction}
The standard cosmological model contains two types of cosmological
perturbations, namely, the curvature perturbations (i.e.~the scalar
mode) and the primordial gravitational waves (GWs) (i.e.~the tensor
mode). Among the two, the existence and the property of the scalar type
perturbations have been established by a number of observations, such as
fluctuations in the cosmic microwave background (CMB) and large scale
structure of the universe
\cite{Bennett:2012zja,Hinshaw:2012aka,Ade:2013zuv,Planck:2013kta,Tegmark:2006az}.
On the other hand, the observational cosmology through the tensor
mode perturbations has just begun with the breaking discovery of
primordial B-mode polarizations of CMB anisotropies by the BICEP2
experiment \cite{Ade:2014xna}. If this B-mode signal is attributed to the
primordial GWs produced during inflation, the scalar-to-tensor ratio is
as large as $r\approx 0.2$.

While the primordial GWs with $r\approx 0.2$ is large
enough to dominate the stochastic background GWs on large scales in the
universe, there exit other cosmological processes that induce
the stochastic background of GWs on small scales (e.g.,
Ref.~\cite{Binetruy:2012ze}). 
For example, first order phase transitions in the early universe create
GWs through collisions of phase bubbles with frequencies corresponding to the Hubble parameter at that epoch
\cite{1986MNRAS.218..629H,PhysRevD.30.272}. The other sources include
exotic matters such as cosmic strings \cite{PhysRevD.45.3447}, inflation
with extra fields such as axion like particles \cite{Sorbo:2011rz}, and
self ordering scalar fields
\cite{JonesSmith:2007ne,Fenu:2009qf,Krauss:2010df}. In these
models, the amplitude of generated GWs strongly depends on the model
setup or its model.

The second-order GWs generated through non-linear couplings with first
order density perturbations is also one of the components of the stochastic
background of GWs on small scales. This GWs production mechanism has a
unique advantage that the amplitude and the spectrum of GWs are
completely predictable.
In extending the cosmological perturbation theory to second-order,
no 
extra model parameter is needed since the statistical properties such as
the amplitude and the spectrum of first-order cosmological perturbations
are already well determined in both theoretical and observational
respects. Because the first order quantities of cosmological perturbations
are precisely determined from the measurements of CMB anisotropies on
large scales by
WMAP and PLANCK experiments, the evolution of perturbations at 
second-order should be determined without ambiguity. Owing to this
advantage, the second-order GWs can be used to
investigate the amplitude of density perturbations on small scales and
the thermal history of the early universe
\cite{2011PhRvD..83h3521B,2010PThPh.123..867S,2009PhRvL.102p1101S,2009PhRvD..79h3511A,2012JCAP...09..017A},
where and when the standard CMB and large scale structure observations
can not reach.
The second-order
cosmological perturbation theory is currently under development and partially
has been established. In
Refs.~\cite{Hu:1993tc,Senatore:2008vi,Bartolo:2007ax,Bartolo:2005kv,Bartolo:2006cu,Bartolo:2006fj,Pitrou:2010sn,Pitrou:2008hy,Beneke:2011kc,Beneke:2010eg},
the Boltzmann equation is expanded up to second-order in cosmological
perturbations including an interaction between photons and electrons.
More recently, the gauge dependence and the invariance are
analyzed in Ref.~\cite{Naruko:2013aaa}.

The worth of the second-order cosmological perturbation theory is not
only in the improvement of accuracy of the theory but also in the
appearance of new effects which do not arise in the first-order
cosmological perturbation theory.
For example, non-Gaussianity arises from non-linearity at 
second-order since the non-linear couplings cause the strain of the
primordial Gaussian profile of perturbations. In the PLANCK experiment, the non-Gaussianity is
characterized by $f_{\rm NL}$, which is found to be consistent with zero
as $f^{\rm local}_{\rm NL} = 2.7\pm 5.8$ \cite{Ade:2013ydc}. When
deriving this constraint, secondary non-Gaussianities have been
estimated and removed 
to pick out the primordial non-Gaussianity, which is a excellent tracer of
inflation, from observed non-Gaussianity
\cite{Bartolo:2010qu,Nitta:2009jp,Pettinari:2013he,Pitrou:2010sn}.
Other effects arising at the second-order include the gravitational
lensing effect, the secondary Doppler effect \cite{1986ApJ...306L..51O},
generation of magnetic fields (vector mode)
\cite{Ichiki:2011ah,Takahashi:2005nd,Fenu:2010kh}, and so on.

In this paper, we especially focus on the secondary GWs. When we expand the
cosmological perturbations up to second-order, the scalar, vector
and tensor modes are no longer independent. Because of the mode
coupling between scalar and tensor perturbations, the product of two first-order scalar
perturbations can induce one tensor perturbation at second-order,
which is equivalent to GWs. In previous works, the mode converting
effect from two scalar modes to one tensor mode has already been studied
\cite{Ananda:2006af,Assadullahi:2009jc,Baumann:2007zm}. However, these
studies have not considered the effect of purely second-order anisotropic
stress of photons and neutrinos. The estimation of the secondary
GWs has not been completed in this sense. 
In Ref.~\cite{Mangilli:2008bw}, they derived the analytic formulae of
second-order anisotropic stresses of photons and neutrinos but did not
discuss effects of the anisotropic stresses on the secondary induced
GWs.
We should note that the second-order Boltzmann codes named \textsf{SONG}
\cite{Pettinari:2013he,Fidler:2014oda} and CosmoLib$2^{\rm nd}$
\cite{PhysRevLett.110.101303} take into account the effects of purely
second-order anisotropic stress on the second-order CMB bispectrum and
the second-order B-mode polarization spectrum.  
Therefore, this paper for the first time analyses the full second-order
gravitational wave spectrum, while its indirect effect has already been
included in the photon bispectrum analysis and in the analysis of the
second-order B-modes in the above-mentioned works.

In the first-order cosmological perturbation theory, anisotropic stress of neutrinos is shown to affect the spectrum of background GWs from inflation at several tens of percent
level
\cite{Pritchard:2004qp,Weinberg:2003ur,Bashinsky:2005tv,Dicus:2005rh,Ichiki:2006rn}.
These studies found that the anisotropic stress pulls out
the energy of GWs, which causes the damping of
GWs and also CMB anisotropies generated from the tensor mode. In
preparation for the future experiments to detect cosmological GWs
directly, such as DECIGO
\cite{Seto:2001qf} and atomic gravitational wave interferometric sensors \cite{PhysRevD.78.122002}, the precise
estimation of the amplitude of GWs would be needed. 
In this paper, we estimate the amplitude of the secondary GWs including 
not only the product of the first-order perturbations but also the purely second-order perturbations. 
To achieve this, we solve the full
system of the Einstein-Boltzmann equations at second-order in the tensor
mode numerically. 

The plan of this paper is as follows. In the next section, we expand
the Einstein and Boltzmann equations up to second-order in the
cosmological perturbation theory.
In section III, we show and discuss results of our numerical
calculation. To set up the numerical calculation, we derive the
solution up to the first-order in the tight-coupling approximation,
which is adopted as our initial condition. 
In section IV, we discuss the effects of second-order anisotropic stress on the gravitational wave spectrum.
Section V is devoted to our summary. 

Throughout this paper, we use the units in which $c=\hbar =1$ and the metric signature with $(-,+,+,+)$.
We obey the rule that the Greek indices run from $0$ to $3$ and the alphabets run from $1$ to $3$, respectively.

\section{second-order perturbation theory}
In this section, we formulate the second-order cosmological perturbation theory.
Throughout this paper, we work in the Poisson gauge \cite{Bartolo:2006cu} whose metric is given by
\begin{equation}
ds^{2}=a^{2}(\eta)\left[ -e^{2\Psi}d\eta^{2}+2\omega_{i}d\eta dx^{i}+\left( e^{-2\Phi}\delta_{ij}+\chi_{ij}\right) dx^{i}dx^{j}\right] ~,
\end{equation}
where the gauge conditions $\omega^{i}{}_{,i}=\chi^{ij}{}_{, j}=0$ and the traceless condition $\chi^{i}{}_{i} = 0$ are imposed on $\omega_{i}$ and $\chi_{ij}$.
Raising or lowering indices of perturbations are done by $\delta_{ij}$.
Owing to these gauge conditions and the traceless condition,
$\omega_{i}$ and $\chi_{ij}$ contain only the vector and tensor modes, respectively.

In the second-order cosmological perturbation theory, scalar, vector and
tensor modes must mix due to the non-linearity. Scalar, vector and
tensor modes correspond to the perturbations of density or curvature,
vorticity, and GWs,
respectively. 
We neglect the first-order vector mode since the vector mode has only a decaying solution in the standard cosmology. We also neglect the first order
tensor mode for clarity because our aim here is to precisely estimate the amplitude
of the second-order tensor perturbations which are generated from the first order density perturbations.
We
expand cosmological perturbations in the metric as $\Psi =
\Psi^{(1)}+\frac{1}{2}\Psi^{(2)}$, $\Phi =
\Phi^{(1)}+\frac{1}{2}\Phi^{(2)}$, $\omega_{i} =
\frac{1}{2}\omega^{(2)}_{i}$, and $\chi_{ij} =
\frac{1}{2}\chi^{(2)}_{ij}$. 

In Ref.~\cite{Baumann:2007zm},
the authors only
considered the secondary GWs induced through the convolutions
of the first-order scalar perturbations, in which case the second-order
Boltzmann equation is not necessary.
In this paper, we need solve the second-order Boltzmann
equation because we consider not only the convolutions but also purely second-order effects.

\subsection{Boltzmann equation}
The Boltzmann equation describes the evolution of distribution functions of
particles including microscopic collisions.
Let us consider the Boltzmann equation for photons, which is given by
\begin{equation}
\frac{df}{d\lambda}(x^{\mu}, P^{\mu})=\tilde{C}\left[ f\right]~,
\end{equation}
where $f(x^{\mu}, P^{\mu})$ is the distribution function, $P^{\mu}$ is the canonical momentum, and $\lambda$ is the affine parameter and $\tilde{C}\left[ f\right]$
is the collision term due to the Thomson interaction between photons and 
electrons. Here we omit the interaction between photons and
protons because the Thomson cross section of protons is much smaller than that of
electrons.
In the Boltzmann equation for dark
matter or neutrinos, the collision term must vanish.

To calculate the perturbed Boltzmann equation, it is useful to change the coordinate system from Poisson gauge $(x^{\mu},P^{\mu})$ to the local inertial frame $(x^{\mu}, p^{\mu})$ \cite{Senatore:2008vi}.
Since we consider the cosmological perturbations up to
second-order, the distribution function is expanded as 
\begin{equation}
f(\eta, \bm{x},p,\hat{n})=f^{(0)}(\eta, p) + f^{(1)}(\eta, \bm{x},p,\hat{n})+\frac{1}{2}f^{(2)}(\eta, \bm{x},p,\hat{n})~,
\end{equation}
where $p$ and $\hat{n}$ are the amplitude and the direction of
photon's momentum, respectively.
The zeroth order distribution function, $f^{(0)}(\eta, p)$, is fixed
to the Planck distribution.
It is useful to define the brightness function which is given by
\begin{equation}
\Delta^{(1,2)} (\eta, \bm{x}, \hat{n})=\frac{\int{dp~p^{3}f^{(1,2)}}(\eta , \bm{x}, p, \hat{n})}{\int{dp~p^{3}f^{(0)}}(\eta , p)}~,
\end{equation}
where the denominator of the right-hand side is proportional to the
mean energy density of photons. 
The relations between the temperature fluctuation of CMB, $\Theta\equiv \delta T/T$, and the brightness function are given by $\Delta^{(1)}= 4\Theta^{(1)}$ and $\Delta^{(2)}=4\Theta^{(2)}+16( \Theta^{(1)})^{2}$ \cite{Nitta:2009jp} at first and second order, respectively.

The angle dependence of the brightness function is expanded by the
spherical harmonics as 
\begin{equation}
\Delta^{(1,2)}(\eta, \bm{x},\hat{n})=\sum_{\ell}\sum^{\ell}_{m=-\ell}\Delta^{(1,2)}_{\ell ,m}(\eta, \bm{x})(-i)^{\ell}\sqrt{\frac{4\pi}{2\ell +1}}Y_{\ell ,m}(\hat{n})~.
\end{equation}
The Boltzmann equation of photons in terms of $\Delta^{(2)}_{\ell, m}$
up to second-order is written by 
\begin{equation}
\dot{\Delta}^{(2)}_{\ell ,m}+k\left[ \frac{c_{\ell +1, m}}{2\ell +3}\Delta^{(2)}_{\ell +1, m}-\frac{c_{\ell ,m}}{2\ell -1}\Delta^{(2)}_{\ell -1, m}\right]
=S^{(2)}_{\ell ,m}~,
\end{equation}
where $c_{\ell, m}\equiv \sqrt{\ell^{2}-m^{2}}$.
A dot represents a derivative with respect to the conformal time $\eta$.
Here we have translated from real space to Fourier space, following
the convention of the Fourier transformation as
\begin{equation}
f(\bm{x}) = \int{\frac{d^{3}k}{(2\pi)^{3}} f(\bm{k})}e^{i\bm{k}\cdot \bm{x}} ~.
\end{equation}

The source term $S^{(2)}_{\ell ,m}$ can be expressed as
\begin{equation}
S^{(2)}_{\ell ,m}(\bm{k},\eta )=\mathcal{C}^{(2)}_{\ell, m}(\bm{k},\eta ) +\mathcal{G}^{(2)}_{\ell, m}(\bm{k},\eta ) ~.\label{scat term}
\end{equation}
Here $\mathcal{C}^{(2)}_{\ell, m}$ is the collision term that is
proportional to the differential optical depth $\dot{\tau}_{c}\equiv -an_{e}\sigma_{\rm T}$, where $a$, $n_{e}$, and $\sigma_{\rm T}$
are the number density of the electron, scale factor, and the Thomson scattering cross-section, respectively, and $\mathcal{G}^{(2)}_{\ell, m}$
denotes the gravitational effects, i.e., the lensing and the redshift
terms. In this paper, we call $\mathcal{C}_{\ell, m}$ and
$\mathcal{G}_{\ell ,m}$ the scattering term and the gravitational term,
respectively. The explicit form of $\mathcal{C}^{(2)}_{\ell, m}$ and
$\mathcal{G}^{(2)}_{\ell, m}$ are given as
\begin{widetext}
\begin{eqnarray}
\mathcal{C}^{(2)}_{\ell, m}
&=&\dot{\tau}_{c}\Delta^{(2)}_{\ell ,m}-\dot{\tau}_{c}\left( \Delta^{(2)}_{00}\delta_{\ell ,0}\delta_{m, 0}+4v^{(2)}_{{\rm b}~m}\delta_{\ell ,1}+\frac{1}{10}\Delta^{(2)}_{2, m}\delta_{\ell ,2}\right) \nonumber \\
&& +\dot{\tau}_{c}\int{\frac{d^{3}k_{1}}{(2\pi)^{3}}}\left[ -2(\delta_{\rm b}+\Psi)^{(1)}(k_{1}) \delta^{(1)}_{\gamma}(k_{2})-4(\hat{k}_{1}\cdot\hat{k}_{2})v^{(1)}_{\gamma~0}(k_{1})v^{(1)}_{{\rm b}~0}(k_{2}) \right] \delta_{\ell ,0}\delta_{m ,0} \nonumber \\
&&+\dot{\tau}_{c}\int{\frac{d^{3}k_{1}}{(2\pi)^{3}}}\left[ -2(\hat{k}_{1}\cdot\hat{k}_{2}) v^{(1)}_{{\rm b}~0}(k_{1})v^{(1)}_{{\rm b}~0}(k_{2})\right]\delta_{\ell ,0}\delta_{m,0} \nonumber \\
&&+\dot{\tau}_{c}\int{\frac{d^{3}k_{1}}{(2\pi)^{3}}}\left[ -8v^{(1)}_{{\rm b}~0}(k_{1})(\delta_{\rm b}+\Psi)^{(1)}(k_{2}) -6v^{(1)}_{{\rm b}~0}(k_{1})\delta^{(1)}_{\gamma}(k_{2}) -2v^{(1)}_{{\rm b}~0}(k_{1})\Pi^{(1)}_{\gamma~0} (k_{2}) \right] \sqrt{\frac{4\pi}{3}}Y^{*}_{1 ,m}(\hat{k}_{1}) \delta_{\ell ,1}\nonumber \\
&&+\dot{\tau}_{c}\int{\frac{d^{3}k_{1}}{(2\pi)^{3}}}\left[ -\Pi^{(1)}_{\gamma~0}(k_{1}) (\delta_{\rm b}+\Psi)^{(1)}(k_{2}) \right] \sqrt{\frac{4\pi}{5}}Y^{*}_{2,m}(\hat{k}_{1})\delta_{\ell ,2} \nonumber \\
&&+\dot{\tau}_{c}\int{\frac{d^{3}k_{1}}{(2\pi)^{3}}}\left[ 2\Delta^{(1)}_{\ell ,0}(k_{1})(\delta_{\rm b}+\Psi)^{(1)}(k_{2}) \right] \sqrt{\frac{4\pi}{2\ell +1}}Y^{*}_{\ell ,m}(\hat{k}_{1}) \nonumber \\
&&+\dot{\tau}_{c}i(-i)^{-\ell}(-1)^{m}(2\ell +1)\sum_{\ell_{1}}\sum_{m_{1},m_{2}}(-i)^{\ell_{1}}
\left(
\begin{array}{ccc}
\ell_{1} & 1 & \ell \\
0 & 0 & 0
\end{array}
\right) 
\left(
\begin{array}{ccc}
\ell_{1} & 1 & \ell \\
m_{1} & m_{2} & -m
\end{array}
\right) \nonumber \\
&&\times \int\frac{d^{3}k_{1}}{(2\pi)^{3}} \left[ (2+\delta_{\ell_{1} ,2})\Delta^{(1)}_{\ell_{1} ,0}(k_{1})v^{(1)}_{{\rm b}~0}(k_{2}) \right] \sqrt{\frac{4\pi}{2\ell_{1} +1}}Y^{*}_{\ell_{1} ,m_{1}}(\hat{k}_{1})\sqrt{\frac{4\pi}{3}}Y^{*}_{1, m_{2}}(\hat{k}_{2}) \nonumber \\
&&+\dot{\tau}_{c}(-i)^{-\ell}(-1)^{m}(2\ell +1)\sum_{m_{1} ,m_{2}}
\left(
\begin{array}{ccc}
1 & 1 & \ell \\
0 & 0 & 0
\end{array}
\right) 
\left(
\begin{array}{ccc}
1 & 1 & \ell \\
m_{1} & m_{2} & -m
\end{array}
\right) \nonumber \\
&&\times 
\int{\frac{d^{3}k_{1}}{(2\pi)^{3}}}\left[ 14v^{(1)}_{{\rm b}~0}(k_{1})v^{(1)}_{{\rm b}~0}(k_{2}) -4v^{(1)}_{\gamma~0}(k_{1})v^{(1)}_{{\rm b}~0}(k_{2}) \right]\sqrt{\frac{4\pi}{3}}Y^{*}_{1 ,m_{1}}(\hat{k}_{1})\sqrt{\frac{4\pi}{3}}Y^{*}_{1 ,m_{2}}(\hat{k}_{2}) ~, 
\end{eqnarray}
and
\begin{eqnarray}
\mathcal{G}^{(2)}_{\ell, m}
&=&4\dot{\Phi}^{(2)}\delta_{\ell ,0}\delta_{m, 0}
-\sum_{\lambda =\pm 1}4\dot{\omega}^{(2)}_{\lambda}\delta_{\ell ,1}\delta_{m ,\lambda}+4k\Psi^{(2)}\delta_{\ell ,1}\delta_{m,0}-\sum_{\sigma = \pm 2}2\dot{\chi}^{(2)}_{\sigma}\delta_{\ell ,2}\delta_{m,\sigma} \nonumber \\
&&+4k\int{\frac{d^{3}k_{1}}{(2\pi)^{3}}}\left[ \Psi^{(1)}(k_{1})\Psi^{(1)}(k_{2}) \right] \delta_{\ell ,1}\delta_{m,0} +\int{\frac{d^{3}k_{1}}{(2\pi)^{3}}}\left[ 8\Delta^{(1)}_{\ell ,0}(k_{1})\dot{\Phi}^{(1)}(k_{2}) \right] \sqrt{\frac{4\pi}{2\ell +1}}Y^{*}_{\ell ,m}(\hat{k}_{1}) \nonumber \\
&&+2i(-i)^{-\ell}(2\ell +1)\sum_{\ell_{2}}\sum_{m_{1},m_{2}} (-1)^{m}(2\ell_{2}+1)
\left(
\begin{array}{ccc}
1 & \ell_{2} & \ell \\
0 & 0 & 0
\end{array}
\right) 
\left(
\begin{array}{ccc}
1 & \ell_{2} & \ell \\
m_{1} & m_{2} & -m
\end{array}
\right) \nonumber \\
&&\times \int{\frac{d^{3}k_{1}}{(2\pi)^{3}}}\left[ k_{1}\left( \Psi+\Phi\right)^{(1)}(k_{1}) \tilde{\Delta}^{(1)}_{\ell}(k_{2}) \right] \sqrt{\frac{4\pi}{3}}Y^{*}_{1,m_{1}}(\hat{k}_{1})\sqrt{\frac{4\pi}{2\ell_{2}+1}}Y^{*}_{\ell_{2},m_{2}}(\hat{k}_{2}) \nonumber \\
&&+\int{\frac{d^{3}k_{1}}{(2\pi)^{3}}}\left[ 8k_{1}\Psi^{(1)}(k_{1})\Phi^{(1)}(k_{2}) \right] \sqrt{\frac{4\pi}{3}}Y^{*}_{1 ,m}(\hat{k}_{1}) \delta_{\ell ,1} \nonumber \\
&&-i(-i)^{-\ell}(2\ell +1)\sum_{\ell_{1}}(-i)^{\ell_{1}}
\left(
\begin{array}{ccc}
\ell_{1} & 1 & \ell \\
0 & 0 & 0
\end{array}
\right) 
\left(
\begin{array}{ccc}
\ell_{1} & 1 & \ell \\
0 & 0 & 0
\end{array}
\right) 
\int\frac{d^{3}k_{1}}{(2\pi)^{3}} \left[ 2k_{1}\Delta^{(1)}_{\ell_{1} ,0}(k_{1})(\Psi^{(1)}+\Phi^{(1)})(k_{2}) \right] \sqrt{\frac{4\pi}{2\ell+1}}Y^{*}_{\ell,m}(\hat{k}_{1}) \nonumber \\
&&-i(-i)^{-\ell}(-1)^{m}(2\ell +1)\sum_{\ell_{1}}\sum_{m_{1}, m_{2}}(-i)^{\ell_{1}}
\left(
\begin{array}{ccc}
\ell_{1} & 1 & \ell \\
0 & 0 & 0
\end{array}
\right) 
\left(
\begin{array}{ccc}
\ell_{1} & 1 & \ell \\
m_{1} & m_{2} & -m
\end{array}
\right) \nonumber \\
&&\times\int\frac{d^{3}k_{1}}{(2\pi)^{3}} \left[ 8k_{2}\Delta^{(1)}_{\ell_{1} ,0}(k_{1}) \Psi^{(1)}(k_{2}) \right] \sqrt{\frac{4\pi}{2\ell_{1} +1}}Y^{*}_{\ell_{1} ,m_{1}}(\hat{k}_{1})\sqrt{\frac{4\pi}{3}}Y^{*}_{1, m_{2}}(\hat{k}_{2}) \nonumber \\
&&+2i(-i)^{\ell}(-1)^{m}(2\ell +1)\sum_{L,L',L''}\sum_{M',M''}(2L+1)(2L'+1)(2L''+1) \nonumber \\
&&\times
\left(
\begin{array}{ccc}
1 & 1 & L' \\
0 & 0 & 0
\end{array}
\right)
\left(
\begin{array}{ccc}
L' & \ell & L \\
0 & 0 & 0
\end{array}
\right)
\left(
\begin{array}{ccc}
1 & L & L'' \\
0 & 0 & 0
\end{array}
\right)
\left(
\begin{array}{ccc}
1 & L'' & \ell \\
M' & M'' & -m
\end{array}
\right)
\left\{
\begin{array}{ccc}
1 & \ell & L'' \\
L & 1 & L'
\end{array}
\right\} \nonumber \\
&&\times \int{\frac{d^{3}k_{1}}{(2\pi)^{3}}}\left[ k_{1}\left( \Psi+\Phi\right)^{(1)}(k_{1})\tilde{\Delta}^{(1)}_{L}(k_{2})\right] \sqrt{\frac{4\pi}{3}}Y^{*}_{1,M'}(\hat{k}_{1})\sqrt{\frac{4\pi}{2L''+1}}Y^{*}_{L'',M''}(\hat{k}_{2}) ~, \label{gravity boltz 2nd}
\end{eqnarray}
\end{widetext}
where Fourier wavevectors $\bm{k}$, $\bm{k}_{1}$, and $\bm{k}_{2}$ satisfy the relation $\bm{k} = \bm{k}_{1}+\bm{k}_{2}$. The relations between the distribution function and the density perturbation $\delta$, the velocity perturbation $v$, and anisotropic stress $\Pi_{\gamma}$ are defined in Ref.~\cite{Bartolo:2010qu}.
In Eq.~(\ref{gravity boltz 2nd}), we have defined
$\tilde{\Delta}^{(1)}_{\ell}$ as
\begin{equation}
\tilde{\Delta}^{(1)}_{\ell''} \equiv (2\ell'' +3)\Delta^{(1)}_{\ell''+1}+(2\ell'' +7)\Delta^{(1)}_{\ell''+3}+\cdots ~ , \label{def lens term}
\end{equation}
which comes from the lensing term \cite{Nitta:2009jp}.
We see that the lensing term contains higher multipole moments.
The source term of the first-order Boltzmann equation vanish
when $m\neq 0$, because we consider only the scalar mode in the first
order perturbations. 
However for the second-order perturbations, 
not only the
scalar mode $(m=0)$, but also the vector $(m=\lambda)$ and tensor
$(m=\sigma)$ modes arise due to non-linear couplings, where $\lambda = \pm 1$ and $\sigma = \pm 2$, respectively.
Note that in the Einstein gravity, there is no source of the modes with $|m|\geq 3$.

When considering massless neutrinos, one can set $\dot{\tau}_{c}=0$
in the above equations because massless neutrinos interact with the other fluids only
through gravity. We do not write down the hierarchical equation of
neutrinos here since it is trivial. The distribution function of
neutrinos is also expanded by the spherical harmonics and we write the
expansion coefficients as $\mathcal{N}^{(1,2)}_{\ell, m}$ in this paper.

\subsection{Tensor decomposition of the Einstein equation}
Let us write down the second-order Einstein equation.
Here we concentrate only on the tensor mode, which is equivalent to GWs.
The second-order Einstein and energy-momentum tensors are, respectively~\cite{Bartolo:2010qu},
\begin{eqnarray}
a^{2}G^{i}{}_{j} &=&
e^{2\Phi}\left( \Phi^{,i}{}_{,j}-\Psi^{,i}{}_{,j}\right) + \Phi^{,i}\Phi_{,j}-\Psi^{,i}\Psi_{,j}-\left( \Phi^{,i}\Psi_{,j}+\Phi_{,j}\Psi^{,i}\right) \notag \\
&&+\mathcal{H}\left[ \dot{\chi}^{i}{}_{j}-\left( \omega^{i}{}_{,j}+\omega_{j}{}^{,i}\right)\right] +\frac{1}{2}\left[ \ddot{\chi}^{i}{}_{j}-\left( \dot{\omega}^{i}{}_{,j}+\dot{\omega}_{j}{}^{,i}\right)-\chi^{i}{}_{j}{}^{,a}{}_{,a} \right] \notag \\
&&+ \mbox{(diagonal part)}~\delta^{i}{}_{j}~, \label{Einstein t}
\end{eqnarray}
and 
\begin{eqnarray}
T_{\rm r}^{i}{}_{j}&=&\rho_{\rm r}\Pi_{\rm r}^{i}{}_{j}+\mbox{(diagonal part)}~\delta^{i}{}_{j} ~, \\
T_{\rm m}^{i}{}_{j}&=&\rho_{\rm m}v^{(1)}_{{\rm m} i}v^{(1)}_{{\rm m} j}+\mbox{(diagonal part)}~\delta^{i}{}_{j} ~, \label{em t}
\end{eqnarray}
where $T_{\rm r}^{i}{}_{j}$ and $T_{\rm m}^{i}{}_{j}$ denote massless
(relativistic) particles such as photons and neutrinos, and massive
(non-relativistic) particles such as baryons and dark matter, respectively.
As the GWs are equivalent to the traceless and
transverse part of the metric perturbations, we do not pick up the
diagonal part in the Einstein and the energy momentum tensors for
the non-relativistic matters.
We decompose the tensor mode by the following expansion,
\begin{equation}
\chi^{(2)}_{ij}(\bm{x}, \eta) = \int{\frac{d^{3}k}{(2\pi)^{3}}}e^{i\bm{k}\cdot\bm{x}}\sum_{\sigma = \pm 2} \chi^{(2)}_{\sigma}(k,\eta)e^{(\sigma)}_{ij}(\hat{k}) ~,
\end{equation}
where $e^{(\sigma)}_{ij}(\hat{k})$ is the polarization tensor and is constructed by the polarization vectors as
\begin{equation}
e^{(\pm 2)}_{ij}(\hat{k}) = -\sqrt{\frac{3}{2}}\epsilon^{(\pm 1)}_{i}(\hat{k})\epsilon^{(\pm 1)}_{j}(\hat{k})~.
\end{equation}
This polarization tensor satisfies the traceless and transverse conditions as
\begin{equation}
\hat{k}^{i}e^{(\pm 2)}_{ij}(\hat{k}) = e^{(\pm 2) i}{}_{i}(\hat{k}) = 0 ~.
\end{equation}
By contracting Eqs.~(\ref{Einstein t}) and~(\ref{em t}) with $e^{(-\sigma)}_{ij}$, we can obtain the tensor part of the Einstein equation as
\begin{widetext}
\begin{eqnarray}
\ddot{\chi}^{(2)}_{\sigma}+2\mathcal{H}\dot{\chi}^{(2)}_{\sigma}+k^{2}\chi^{(2)}_{\sigma}
&=&8\pi Ga^{2}\rho^{(0)}_{\gamma} \frac{4}{15}\Delta^{(2)}_{2,\sigma} +8\pi Ga^{2}\rho^{(0)}_{\nu} \frac{4}{15}\mathcal{N}^{(2)}_{2,\sigma} \notag \\
&&+\sum_{{\rm s}= {\rm b}, {\rm dm}}8\pi Ga^{2}\rho^{(0)}_{\rm s}\int{\frac{d^{3}k_{1}}{(2\pi)^{3}}} \left[ 4\sqrt{\frac{2}{3}}v^{(1)}_{{\rm s}~0}(k_{1})v^{(1)}_{{\rm s}~0}(k_{2}) \right] \sqrt{\frac{4\pi}{3}}Y^{*}_{1,\lambda}(\hat{k}_{1})\sqrt{\frac{4\pi}{3}}Y^{*}_{1,\lambda}(\hat{k}_{2}) \notag \\
&+&\int\frac{d^{3}k_{1}}{(2\pi)^{3}}\frac{8}{3}k^{2}_{1}\left[ \Phi^{(1)}(k_{1})\Phi^{(1)}(k_{2})+\Psi^{(1)}(k_{1})\Psi^{(1)}(k_{2}) \right] \sqrt{\frac{4\pi}{5}}Y^{*}_{2,\sigma}(\hat{k}_{1}) ~, \label{eq: GWs}
\end{eqnarray}
\end{widetext}
where subscripts of ``$\rm b$'' and ``$\rm dm$'' mean baryons and dark
matter, respectively, $\mathcal{H}=\dot{a}/a$, and
$\Delta^{(2)}_{2,\sigma}$ and $\mathcal{N}^{(2)}_{2, \sigma}$ are
anisotropic stresses of photons and neutrinos, respectively.
The third term of r.h.s. in Eq.~(\ref{eq: GWs}) $\sim v^{(1)}_{{\rm s}}(k_{1})v^{(1)}_{{\rm s}}(k_{2})$ can be read as the anisotropic stress for the non-relativistic matters,
while the products of the velocity perturbations for the relativistic particles are included in the first and second terms.
These purely second-order anisotropic stresses of photons and neutrinos
have not yet been considered in the previous work \cite{Baumann:2007zm}.
Here we newly take into account these contributions to the amplitude
of GWs and find that the effect of the stress is significant, as we shall
show below.

The energy density spectrum of the GWs is defined as (see e.g., \cite{Watanabe:2006qe})
\begin{equation}
\Omega^{(2)}_{\rm GW}\equiv \frac{k^{2}}{6\mathcal{H}^{2}}\left[ \frac{k^{3}}{2\pi^{2}}P^{(2)}_{\chi}(k)\right] ~,
\end{equation}
where $P^{(2)}_{\chi}(k)$ is the spectrum of the second-order tensor perturbations
We present these spectra and their time evolutions in the next section.

\subsection{Structure of the second-order perturbation theory}
In the second-order perturbation theory, the transfer function depends on $(k, k_{1}, k_{2})$ or equivalently $(k, \mu_{1}, k_{1})$, where $\mu_{1}$ is defined as $\hat{k}\cdot \hat{k}_{1}$.
The equation below is a schematic equation for the evolution of
second-order perturbations~\cite{Pettinari:2014vja},
\begin{equation}
\hat{\mathcal{L}}\left[ \Delta^{(2)} (\bm{k}, \eta) \right] = \int{\frac{d^{3}k_{1}}{(2\pi)^{3}}}\left[ \mathcal{S}(\bm{k},\eta ; \bm{k}_{1}, \bm{k}_{2})\right]~,
\label{eq:2nd_Boltz}
\end{equation}
where $\bm{k} = \bm{k}_{1} + \bm{k}_{2}$, $\hat{\mathcal{L}}$ is a general linear operator and $\mathcal{S}$ is a source term which is constructed from the first-order perturbation variables.
The source term $\mathcal{S}(\bm{k},\eta ; \bm{k}_{1}, \bm{k}_{2})$ is given by the products of the first order perturbations, 
which can be expressed using the linear transfer functions $\Delta_{{\rm T}1}(k_{1},\eta)$ and $\Delta_{{\rm T}2}(k_{2},\eta)$,
and primordial curvature perturbations $\psi(\bm{k}_{1})$ and $\psi(\bm{k}_{2})$ as
$\Delta_{{\rm T}1}(k_{1},
\eta) \psi(\bm{k}_{1})\Delta_{{\rm T}2}(k_{2}, \eta) \psi(\bm{k}_{2})$, where we used the fact that linear transfer functions do not depend on the direction of the wavevector.
The statistics of
$\psi(\bm{k})$ obeys the random Gaussian and characterized by the primordial power spectrum, as
\begin{equation}
\Braket{\psi(\bm{k}_{1})\psi^{*}(\bm{k}_{2})} = (2\pi)^{3}P(k_{1})\delta(\bm{k}_{1}-\bm{k}_{2})~,
\end{equation}
where $P(k)$ is the power spectrum.

Observationally, the primordial power spectrum is nearly scale-invariant and parameterized as \cite{Bennett:2012zja,Ade:2013uln} 
\begin{equation}
\frac{k^{3}}{2\pi^{2}}P(k)=\frac{4}{9}\Delta^{2}_{\mathcal{R}}(k_{0})\left( \frac{k}{k_{0}}\right)^{n_{s}-1} ~.
\end{equation}
In this paper, 
we employ the standard cosmological model, i.e., the $\Lambda$-CDM model, and
we set $\Delta^{2}_{\mathcal{R}}(k_{0}=0.002~h{\rm
Mpc}^{-1}) = 2.4\times 10^{-9}$ \cite{Bennett:2012zja} and consider a scale-invariant
scalar spectrum with $n_{s} = 1$ for illustration purposes. 
Because we split perturbations into transfer functions and primordial
perturbations, the second-order evolution equation
(Eq.~\ref{eq:2nd_Boltz}) is translated to the equation for second
order transfer functions in $(\bm{k},k_{1},k_{2})$ space as
\begin{equation}
\hat{\mathcal{L}}\left[ \Delta^{(2)}_{\rm T} (\bm{k}, \eta; k_{1}, k_{2}) \right] = \mathcal{S}(\bm{k},\eta ; k_{1}, k_{2})~,
\end{equation}
where the subscript ${\rm T}$ represents the transfer function.
Taking an ensemble average is the final step to derive the power spectra
of the second-order perturbations. 
We can split any second-order perturbation variable into a transfer function and the first-order primordial perturbations as
\begin{equation}
\Delta^{(2)}(\bm{k},\eta;\bm{k}_{1},\bm{k}_{2})=\Delta^{(2)}_{\rm T}(\bm{k},\eta ;k_{1}, k_{2})\times \psi(\bm{k}_{1})\psi(\bm{k}_{2}) ~.
\end{equation}
From the above expression, we can calculate the spectrum of the second-order variable as
\begin{equation}
\Braket{\Delta^{(2)}(\bm{k},\eta;\bm{k}_{1},\bm{k}_{2})\Delta^{*(2)}(\bm{k'},\eta ;\bm{k'}_{1},\bm{k'}_{2})}
=\left[ \Delta^{(2)}_{\rm T}(\bm{k},\eta;k_{1},k_{2})\right]\left[ \Delta^{(2)}_{\rm T}(\bm{k'},\eta ; k'_{1},k'_{2})\right]\times \Braket{ \psi(\bm{k}_{1})\psi(\bm{k}_{2})\psi^{*}(\bm{k'}_{1})\psi^{*}(\bm{k'}_{2})} ~.
\end{equation}
By using Wick's theorem, the bracket in the above equation is reduced to
\begin{eqnarray}
\Braket{\psi(\bm{k}_{1})\psi(\bm{k}_{2})\psi^{*}(\bm{k'}_{1})\psi^{*}(\bm{k'}_{2})}
&=&\Braket{\psi(\bm{k}_{1})\psi^{*}(\bm{k'}_{1})}\Braket{\psi(\bm{k}_{2})\psi^{*}(\bm{k'}_{2})}+\Braket{\psi(\bm{k}_{1})\psi^{*}(\bm{k'}_{2})}\Braket{\psi(\bm{k}_{2})\psi^{*}(\bm{k'}_{1})} \notag\\
&=&(2\pi)^{6}P(k_{1})P(k_{2})\left[ \delta (\bm{k'}_{1}-\bm{k}_{1})+\delta (\bm{k'}_{1}-\bm{k}_{2})\right] \delta (\bm{k}-\bm{k'}) ~.
\end{eqnarray}
To proceed the derivation of the power spectrum, we define the second-order spectrum as
\begin{equation}
\Braket{\Delta^{(2)}(\bm{k},\eta;\bm{k}_{1},\bm{k}_{2})\Delta^{*(2)}(\bm{k'},\eta ;\bm{k'}_{1},\bm{k'}_{2})}\equiv (2\pi)^{3}P_{\Delta^{(2)}}(\bm{k},\eta ;\bm{k}_{1},\bm{k}_{2},\bm{k'}_{1},\bm{k'}_{2})\delta(\bm{k}-\bm{k'}) ~.
\end{equation}
Finally we calculate the convolution as
\begin{eqnarray}
P_{\Delta^{(2)}}(\bm{k},\eta)
&=&\int{\frac{d^{3}k_{1}}{(2\pi)^{3}}}\int{\frac{d^{3}k'_{1}}{(2\pi)^{3}}}P_{\Delta^{(2)}}(\bm{k},\eta ;\bm{k}_{1},\bm{k}_{2},\bm{k'}_{1},\bm{k'}_{2}) \notag \\
&=&2\int{\frac{d^{3}k_{1}}{(2\pi)^{3}}} \left[ \Delta^{(2)}_{\rm T}(\bm{k},\eta ;k_{1},k_{2})\right]^{2} P(k_{1})P(k_{2}) ~. \label{powerspectrum2}
\end{eqnarray}
In the second equality in Eq.~(\ref{powerspectrum2}), 
we assume that the source term $
\mathcal{S}(\bm{k},\eta ; k_{1}, k_{2})$ is symmetric with respect to the exchange
of $k_{1}$ and $k_{2}$, which means that the transfer function $\Delta^{(2)}_{\rm
T}(\bm{k},\eta;k_{1},k_{2})$ is also symmetric in $k_1$ and $k_2$.

\section{Numerical results}
To solve cosmological perturbations up to second-order, we need the time
evolutions of the transfer functions of the first order perturbations,
$\Phi^{(1)}(k, \eta)$, $\Psi^{(1)}(k,\eta)$, $\Delta^{(1)}_{\ell, m}(k,
\eta)$, and $\mathcal{N}^{(1)}_{\ell , m}(k,\eta)$ in the Poisson gauge. 
We obtain these variables using the CAMB code \cite{Lewis:1999bs}.
In practice, we store the first-order variables in $k$-space, whose range is taken as $[5\times 10^{-5}~h{\rm Mpc}^{-1}, 10^{2}~h{\rm Mpc}^{-1}]$.
We truncate the first-order hierarchies of photon and neutrino Boltzmann equations at $\ell = 30$ and the second-order hierarchies of them at $\ell = 25$.
We checked that the results are stable against these choices.

In our numerical calculation, we store first-order transfer
functions and solve the Einstein-Boltzmann system up to second-order
and sample second-order transfer functions in the
$(k_{1}, k_{2})$ plane with a fixed real $k$. Owing to the triangle
condition about $k_{1}$, $k_{2}$, and $k$, we reduce the sampling area of the $(k_{1},
k_{2})$ plane, namely, we need solve the equations only in the region
where $|k_{1}-k_{2}|\leq k\leq k_{1}+k_{2}$. Furthermore, as we stock the
first-order transfer functions with a logarithmic interval, this
triangle condition is 
effective to reduce the cost of numerical calculation.

\subsection{Initial conditions}
To solve the second-order equations derived above numerically, we
should set up the initial condition of each perturbation
variable. Thus we first solve the equations analytically with $k\eta\ll 1$ and using the tight coupling
approximation, and find the initial condition at sufficiently early time
for our numerical calculation.

Deep in the radiation dominated era, photon and baryon fluids are tightly coupled because the opacity $\dot{\tau}_{c}$ is large \cite{Ichiki:2011ah,Maeda:2008dv,Takahashi:2007ds,Pitrou:2010ai}.
Although the photon and baryon fluids would behave as a single fluid, there is a small difference in motion between photon and baryon fluids.
For this reason, we can expand the perturbation variables using the tight-coupling parameter which is given by
\begin{equation}
\epsilon \equiv \left| \frac{k}{\dot{\tau}_{c}}\right| 
\sim 10^{-2}\left( \frac{k}{1{\rm Mpc^{-1}}}\right)\left( \frac{1+z}{10^{4}}\right)^{-2}\left( \frac{\Omega_{\rm b}h^{2}}{0.02}\right)^{-1} ~,
\end{equation}
where $\Omega_{\rm b}$ is the baryon density normalized by the critical
density and $h\equiv H_{0}/100\; [\rm{km\; s^{-1}\; Mpc^{-1}}]$ is the
normalized Hubble parameter with $H_{0}$ being the Hubble constant.
In what follows we derive the tight-coupling solution up to first
order to set the initial condition of photon and baryon fluids at second order in cosmological perturbations and to
calculate the evolution of perturbations in a numerically stable manner.

We expand the cosmological perturbation variables using the
tight-coupling parameter up to first order as,
\begin{equation}
\Delta^{({\rm CPT} = 1,2)}=\Delta^{({\rm CPT} = 1,2,~{\rm TCA} = \O)}+\Delta^{({\rm CPT} = 1,2,~{\rm TCA} = I )}~,
\end{equation}
where the Arabic number ($1,2$), and the Roman number ($\O$,$I$) represent the order in the cosmological perturbation
theory (CPT) and the tight coupling approximation (TCA), respectively.
Note that the tight-coupling expansion is independent of the order of cosmological perturbations.

Let us now focus on the tensor mode ($m=\sigma$).
It is useful to define the function $\mathcal{Y}^{\ell_{1},\ell_{2}}_{\ell,m}(\hat{k}_{1},\hat{k}_{2})$ as
\begin{widetext}
\begin{equation}
\mathcal{Y}^{\ell_{1},\ell_{2}}_{\ell,m}(\hat{k}_{1},\hat{k}_{2}) \equiv (-1)^{m}(2\ell + 1)\sum_{m_{1},m_{2}}
\left(
\begin{array}{ccc}
\ell_{1} & \ell_{2} & \ell \\
0 & 0 & 0
\end{array}
\right)
\left(
\begin{array}{ccc}
\ell_{1} & \ell_{2} & \ell \\
m_{1} & m_{2} & -m
\end{array}
\right)
\sqrt{\frac{4\pi}{2\ell_{1}+1}}Y^{*}_{\ell_{1},m_{1}}(\hat{k}_{1})
\sqrt{\frac{4\pi}{2\ell_{2}+1}}Y^{*}_{\ell_{2},m_{2}}(\hat{k}_{2}) ~, \label{eq: calY}
\end{equation}
\end{widetext}
where $\ell_{1}+\ell_{2}+\ell $ must be even because of a property of the Wigner-3j symbol.
Note that, for the special case that $\ell_{1}=0$ or $\ell_{2}=0$, the dependence on $\hat{k}_{1}$ or $\hat{k}_{2}$ vanishes as
\begin{equation}
\mathcal{Y}^{\ell_{1},0}_{\ell,m}(\hat{k}_{1},\hat{k}_{2}) = \sqrt{\frac{4\pi}{2\ell+1}}Y^{*}_{\ell,m}(\hat{k}_{1}) \delta_{\ell, \ell_{1}}~, ~~~
\mathcal{Y}^{0,\ell_{2}}_{\ell,m}(\hat{k}_{1},\hat{k}_{2}) = \sqrt{\frac{4\pi}{2\ell+1}}Y^{*}_{\ell,m}(\hat{k}_{2}) \delta_{\ell, \ell_{2}}~.
\end{equation}

Firstly, using the function defined above we obtain the solution
at zeroth order in the tight coupling approximation as, 
\begin{eqnarray}
\Delta^{(2,\O )}_{2,\sigma} &=&20 \int\frac{d^{3}k_{1}}{(2\pi)^{3}} \left[ v^{(1,\O )}_{\gamma~0}(k_{1})v^{(1,\O )}_{\gamma~0}(k_{2}) \right] \mathcal{Y}^{1,1}_{2, \sigma}(\hat{k}_{1},\hat{k}_{2}) ~, \label{CPT2TCA0 pi g} \\
\Delta^{(2,\O )}_{\ell \geq 3,\sigma}&=&0 ~.
\end{eqnarray}
It is interesting that at second-order in cosmological perturbations,
anisotropic stress of photons arises even at zeroth order in the tight
coupling approximation, while it vanishes at first order in cosmological
perturbations. The anisotropic stress of photons in the second-order survives because of
the coupling between velocity perturbations of photons in the scalar mode as is shown in Eq.~(\ref{CPT2TCA0 pi g}). This result is
consistent with Refs.~\cite{Bartolo:2010qu,Mangilli:2008bw}.

Next, we consider the next order in the tight coupling
approximation (CPT$=2$, TCA$=I$). We find the results as
\begin{widetext}
\begin{eqnarray}
\frac{9}{10}\Delta^{(2,I)}_{2,\sigma}&=&\frac{2}{\dot{\tau}_{c}}\dot{\chi}^{(2,\O )}_{\sigma} 
-\int{\frac{d^{3}k_{1}}{(2\pi)^{3}}}\left[ 9\Pi^{(1,I)}_{\gamma~0}(k_{1})(\delta_{\rm b}^{(1,\O )}-\Phi^{(1,\O )})(k_{2}) \right] \sqrt{\frac{4\pi}{5}}Y^{*}_{2 ,\sigma}(\hat{k}_{1}) \nonumber \\
&&+4\int{\frac{d^{3}k_{1}}{(2\pi)^{3}}}\left[ 9v^{(1,\O )}_{\gamma~0}(k_{1})v^{(1,I)}_{\gamma~0}(k_{2}) +8v^{(1,\O )}_{\gamma~0}(k_{1})\delta v^{(1,I)}_{\gamma {\rm b}~0}(k_{2}) \right]\mathcal{Y}^{1,1}_{2,\sigma}(\hat{k}_{1},\hat{k}_{2}) \nonumber \\
&&+ \int{\frac{d^{3}k_{1}}{(2\pi)^{3}}}\left[ \left( \frac{k_{1}}{\dot{\tau}_{c}}\right)\left( 10\delta^{(1, \O)}_{\gamma}(k_{1})-8\Phi^{(1,\O )}\right)(k_{1})v^{(1,\O )}_{\gamma~0}(k_{2}) \right]\mathcal{Y}^{1,1}_{2,\sigma}(\hat{k}_{1},\hat{k}_{2}) ~, \label{TCA1st pig}\\
\Delta^{(2,I)}_{3,\sigma}&=&-\left( \frac{k}{\dot{\tau}_{c}}\right)\frac{\sqrt{5}}{5}\Delta^{(2,\O )}_{2,\sigma}+15\int\frac{d^{3}k_{1}}{(2\pi)^{3}} \left[ \Pi^{(1,I)}_{\gamma~0}(k_{1})v^{(1,\O )}_{\gamma~0}(k_{2}) \right] \mathcal{Y}^{2,1}_{3,\sigma}(\hat{k}_{1},\hat{k}_{2}) ~, \\
\Delta^{(2,I)}_{\ell\geq 4,\sigma}&=&0 ~.
\end{eqnarray}
\end{widetext}
It might be surprising, but at this order, the octupole moment
($\ell = 3$) can survive because of two source terms. 
One is the anisotropic stress of photons at zeroth order
in the tight-coupling approximation, which generates $\ell =3$ moment
through the 
streaming effect in the left-hand side of the Boltzmann equation. 
The other is a convolution of the first order cosmological perturbations. This term comes from the collision term.
These terms do not appear in the first order cosmological perturbation
theory and the result here is a genuine second-order effect. Note
that, higher multipoles than $\ell=3$ are equal to zero because the source is absent.

By using the tight-coupling solution, we can analyze the behavior of 
anisotropic stress of photons in early times.
If we adopt the adiabatic initial condition for the first-order variables \cite{Ma:1995ey}, the velocity perturbation of photons is proportional to $\eta$ in the Poisson gauge.
The anisotropic stress of photons is proportional to $\eta^{2}$ and therefore we find the initial time dependence of the second-order anisotropic stress of photons must be proportional to $\eta^{4}$ at zeroth order in the tight coupling approximation. In the next section, we will show that these analytic estimates are consistent with our numerical calculation.

Before analyzing the second-order GWs, we focus on the evolutions of anisotropic stresses of photons and neutrinos at
second-order, which are the essential sources of the GWs.
To understand the behavior of the power spectra of anisotropic stresses
of photons and neutrinos, let us first investigate the source terms in the
evolution equation of anisotropic stress,
$S^{(2)}_{2,\sigma}(\bm{k},\eta ; k_{1},k_{2})$, in the subsection below.
\subsection{Sources for photon and neutrino anisotropic stresses}\label{sec: triangle}
Let us investigate the sources for anisotropic stress at second-order to
understand its time evolution. First of all, we note two key
points in order to understand properties of the second-order power
spectrum and source terms. 

First, in the second-order tensor-mode,
the sources of the gravitational waves in Eq.~(\ref{eq: GWs}) are suppressed on small scales in the $\Lambda$CDM model.
Therefore, the sources with
wavenumbers $k_{1}, k_{2} \lesssim k$ mainly contribute to the
second-order anisotropic stress. Second, as for the source terms, we note that the
difference between the sources of photon and neutrino anisotropic
stress is only in the scattering term, $\mathcal{C}^{(2)}_{\ell, m}\propto
\dot{\tau}_{c}$, as is shown in Eq.~(\ref{scat term}).
The source of photon anisotropic stress has two contributions from the gravitational $\mathcal{G}^{(2)}_{\ell,m}$ and the
scattering $\mathcal{C}^{(2)}_{\ell, m}$ terms, while that of neutrinos has only the gravitational term. 
\begin{figure}[t]
\begin{center}
\rotatebox{0}{\includegraphics[width=0.4\textwidth]{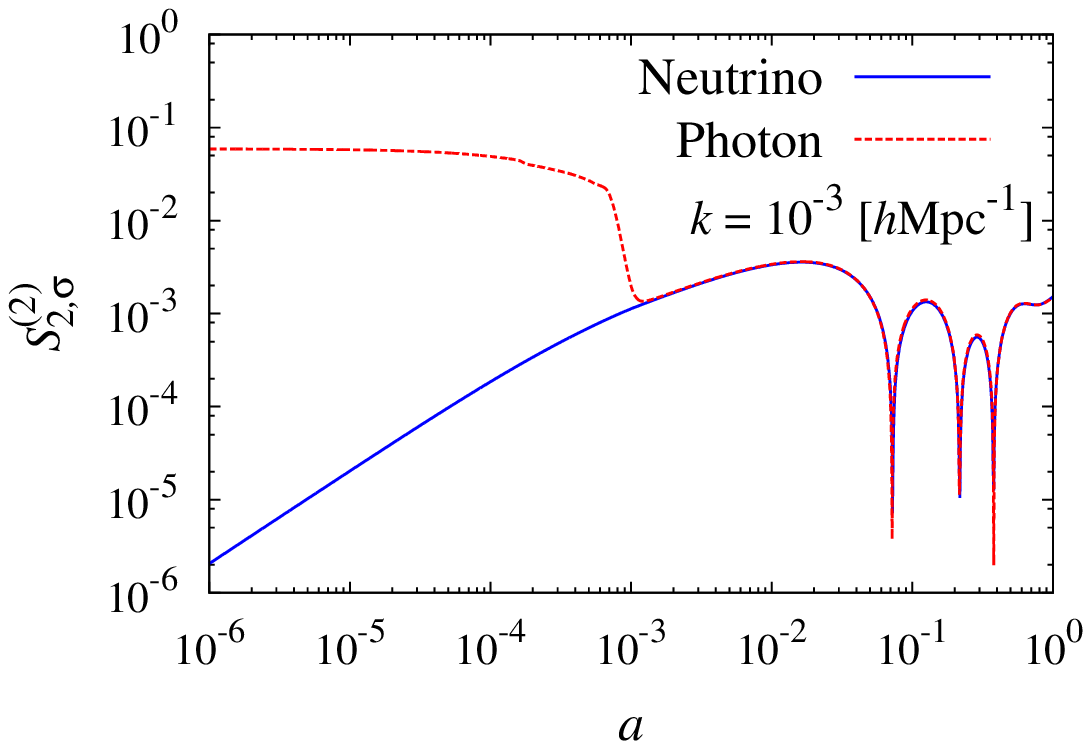}}
\rotatebox{0}{\includegraphics[width=0.4\textwidth]{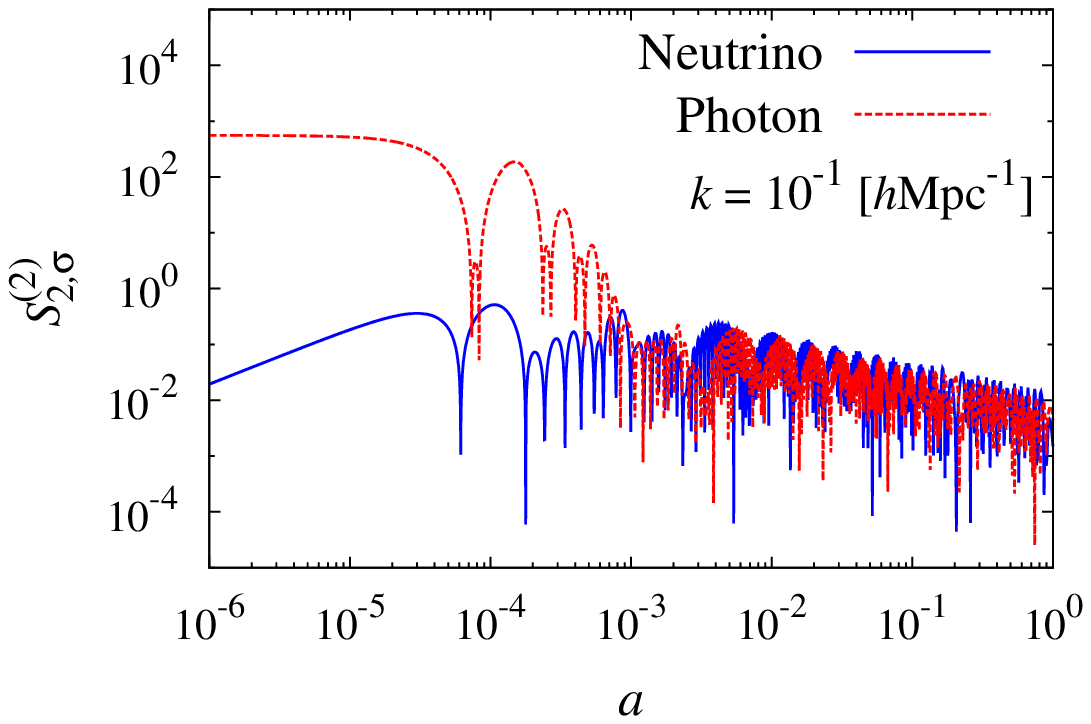}}
\rotatebox{0}{\includegraphics[width=0.4\textwidth]{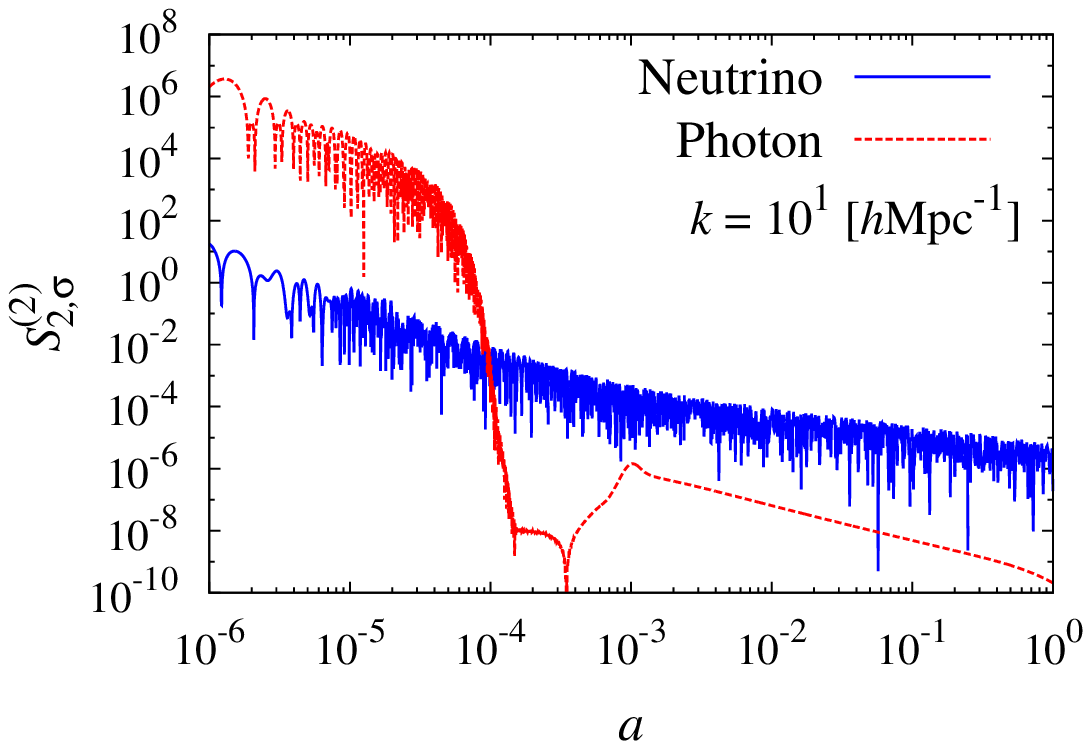}}
\end{center}
\caption{Time evolutions of the source terms of anisotropic stresses of photons and neutrinos $S^{(2)}_{2,\sigma}$ for three typical scales.
Dummy wavenumbers, which are used for convolutions are taken as $k_{1}\approx k_{2}\approx k$.
Blue lines represent the source term of anisotropic stress of neutrinos and red lines do that of photons.
Top left, top right and bottom panels show evolutions at $k\sim 10^{-3}~h{\rm Mpc}^{-1}$, $10^{-1}~h{\rm Mpc}^{-1}$, and $10^{1}~h{\rm Mpc}^{-1}$, respectively.
Note that the source term of photon anisotropic stress before recombination $\eta <\eta_{\rm rec}$ is not used in our numerical calculations because we use the tight-coupling solution for photon anisotropic stress in this epoch.}
\label{fig:source pho neu}
\end{figure}
In Fig.~\ref{fig:source pho neu}, the evolutions of source terms for
anisotropic stresses of photons and neutrinos that satisfy the triangle
configuration as $k\approx k_{1}\approx k_{2}$ are depicted on several scales.

\subsubsection{Photons}
Let us first investigate the evolution of sources for the photon anisotropic stress on large scales ($k\approx
10^{-3}~h$ Mpc$^{-1}$) shown in the top
left panel of Fig.~\ref{fig:source pho neu}.
Before recombination, the scattering term dominates due to the frequent
Compton scattering, while the term becomes negligible compared to the
gravitational term after recombination.
Therefore we can write the source term
of photon anisotropic stress as
\begin{equation}
S^{(2)}_{2,\sigma} \approx
\left\{ \begin{array}{ll}
\mathcal{C}^{(2)}_{2,\sigma} & (\eta < \eta_{\rm rec}) \\
\mathcal{G}^{(2)}_{2,\sigma} & (\eta_{\rm rec} < \eta ) ~,
\end{array} \right.
\end{equation}
while the source of neutrino anisotropic stress has only the gravitational term.
Therefore, after recombination, the source term corresponds to the evolutions of the free-streaming particles.

On intermediate scales ($k\approx 0.01~h{\rm Mpc}^{-1}$, the top
right panel of Fig.~\ref{fig:source pho neu}), there is a subtle
difference between the source terms of photons and neutrinos even after
recombination. This is because the first order perturbations of photons
and neutrinos on these scales enter the horizon before recombination and
evolve differently. On much smaller scales ($k\approx 10~h{\rm
Mpc}^{-1}$, the bottom panel of Fig.~\ref{fig:source pho neu}), the
evolutions of source terms of photons and neutrinos are significantly
different. The source of neutrino anisotropic stress is larger than
that of photons in late times because the Silk damping effect erases
photon perturbations exponentially before recombination. Meanwhile, the source of
neutrino anisotropic stress decays as $\eta^{-1}$ with oscillation
inside the sound horizon because of the free-streaming effect.

\subsubsection{Neutrinos}
Next, let us investigate the evolution of neutrino anisotropic stress.
We find that the source term of neutrino anisotropic
stress is proportional to $\eta^{1}$ on superhorizon scales, since the
source term contains the combination of the form:
\begin{equation}
S^{(2)}_{2,\sigma; \nu} \ni \mathcal{N}^{(1)}_{\ell \geq 1, 0}\times \Psi^{(1)} ~. \label{source approx}
\end{equation}
On super horizon scales, this source term evolves as $\propto \eta^{1}$, since the time
evolutions of $\mathcal{N}^{(1)}_{\ell\geq 1,0}$ and $\Psi^{(1)}$ are given as
$\mathcal{N}^{(1)}_{1,0}\propto \eta$ and $\Psi^{(1)}\propto \eta^{0}$, respectively.
Note that the gravitational lensing terms such as in Eq.~(\ref{source approx}) transport the first-order
higher multipole moments to the second-order lower ones. On the other
hand, gravitational redshift terms transport the first-order lower
multipole moments to the second-order higher ones as seen in Eq.~(\ref{def lens term}).

The structure of the gravitational term of neutrinos is same as that of photons.
At first order, $\Delta^{(1)}_{\ell,m}$ and
$\mathcal{N}^{(1)}_{\ell, m}$ on such large
scales evolve in the same way after recombination since both of
them undergo only free-streaming in a common gravitational potential.
Therefore, we expect that the difference between photons and neutrinos
must vanish on large scales after recombination even at the second
order, as shown in the figure.
\subsection{Photon and neutrino anisotropic stresses}
\begin{figure}[t]
\begin{center}
\rotatebox{0}{\includegraphics[width=0.4\textwidth]{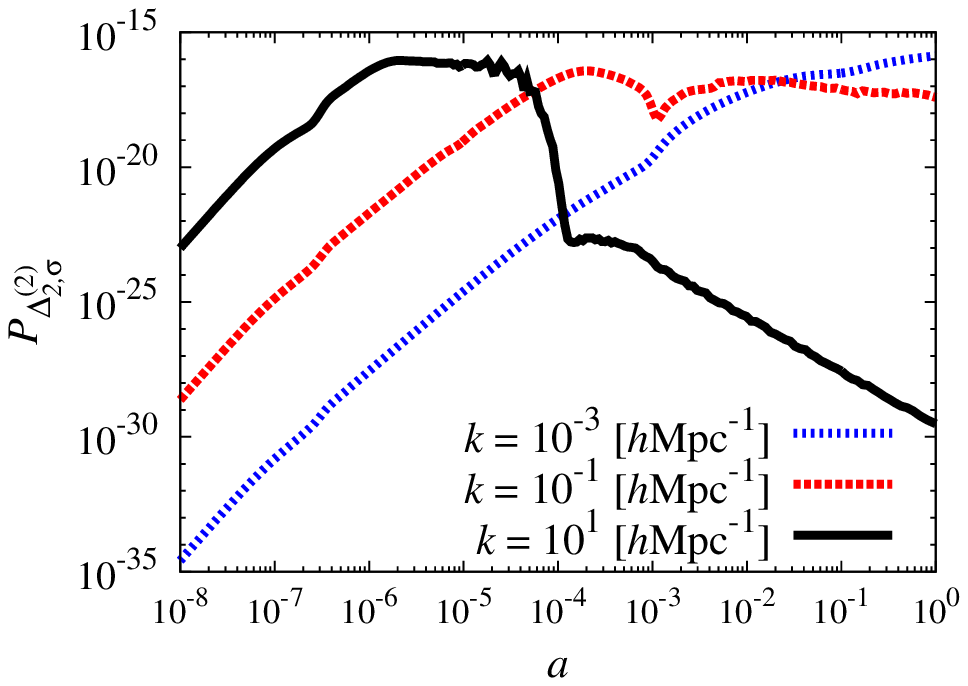}}
\rotatebox{0}{\includegraphics[width=0.4\textwidth]{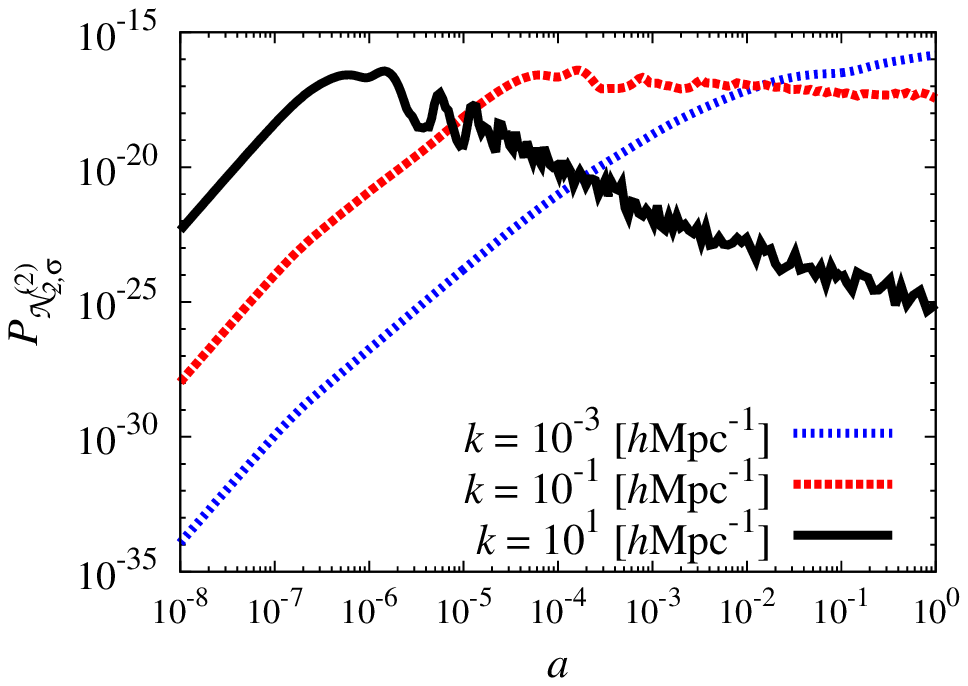}}
\end{center}
\caption{Time evolutions of anisotropic stresses of photons (\textit{left}) and neutrinos (\textit{right}) for wavenumbers $k = 10^{-3}~h{\rm Mpc}^{-1}$, $10^{-1}~h{\rm Mpc}^{-1}$, and $10^{1}~h{\rm Mpc}^{-1}$ as indicated in the panels.
Here we set $\Delta^{2}_{\mathcal{R}}(k_{0}) = 2.4\times 10^{-9}$ and $n_{s} = 1.0$.
It is shown that there are noisy fluctuations due to numerical errors in late time on small scales.
However these fluctuations do not affect on the final result of the spectrum of GWs
since the most of the contribution is coming from the epoch of horizon crossing, e.g., $a\sim 10^{-6}$ for $k = 10^{1}~h{\rm Mpc}^{-1}$.}
\label{fig:evolve pi}
\end{figure}

The evolutions of anisotropic stresses of photons and neutrinos at
second-order, which are the essential sources of the GWs, are
depicted in Fig.~\ref{fig:evolve pi}. 
It is difficult to analyze power spectra of second-order perturbations
quantitatively because the power spectra are obtained after carrying out
complicated convolutions. However we can roughly understand the shape
of the spectra by counting the most contributing terms. 

\subsubsection{Photons}
Let us first
consider the evolution of anisotropic stress of photons on small
scales, in three characteristic epochs separately; (i) the
tight-coupling epoch, (ii) the epoch from when the tight coupling approximation is broken to the onset of recombination, and (iii)
the epoch after recombination.

When the tight-coupling approximation is valid, anisotropic stress
of photons is given by Eq.~(\ref{CPT2TCA0 pi g}) and the power
spectrum of the photon anisotropic stress evolves as $\eta^{4}$, since
Eq.~(\ref{CPT2TCA0 pi g}) indicates that
$\Delta^{(2)}_{2,\sigma}\propto ( v^{(1)}_{\gamma} )^{2}\propto
\eta^{2}$. 
When the zeroth order tight-coupling approximation is broken,
the first-order tight-coupling approximation by Eq.~(\ref{TCA1st pig})
gives a more appropriate
solution for $\Delta^{(2)}_{2,\sigma}$. We find from our numerical
calculation that the right hand side of Eq.~(\ref{TCA1st pig}) is
dominated by the first term given by 
$(2/\dot{\tau_c})\dot{\chi}^{(2)}_{\sigma}$ and also find numerically that $\dot{\chi}^{(2)}_{\sigma}$ evolves as
$\eta^{0.5}$.
Therefore in the first-order tight-coupling approximation, $\Delta^{(2)}_{2,\sigma}\propto \eta^{2.5}$.
The transition from the zeroth order to the first
order solutions can be seen as a kink in the early time evolution of
photon anisotropic stress in the left panel of Fig.~\ref{fig:evolve pi},
for example, at $a\approx 2\times 10^{-7}$ 
for $k=10^{1}~h$Mpc$^{-1}$ (the solid black line).
After the mode enters the horizon, anisotropic stress is sustained by the
source in Eq.~(\ref{TCA1st pig}) until the tight-coupling approximation is broken.

After the tight coupling approximation is broken but before
recombination, the source for the anisotropic stress of photons
experiences the Silk damping effect on small scales. Consequently, the second-order anisotropic
stress of photons loses its source and also experiences the Silk damping.
Note that on the larger scales the first order perturbations do not
experience the Silk damping effect and can sustain the anisotropic
stress at second-order.

After the recombination epoch, the second-order anisotropic stress of photons on
scales smaller than the Silk damping scale
undergoes the free-streaming, since sources of anisotropic stress of
photons have already decayed away.
However, on larger scales, the source of anisotropic stress can survive
owing to the gravitational term (see the modes with $k\lesssim 0.1~h
{\rm Mpc}^{-1}$ in the 
left panel of Fig.~\ref{fig:evolve pi}). 

\subsubsection{Neutrinos}
Next, let us consider the neutrino anisotropic stress, which is sourced only
by the gravitational term. 
Before the horizon crossing, neutrino anisotropic stress is
generated by the source, which is of the form
$S^{(2)}_{2,\sigma; \nu} \ni
\mathcal{N}^{(1)}_{\ell \geq 1, 0}\times \Psi^{(1)}\propto
\eta^{1}\times \eta^{0}$ (see Eq.~(10)).
This means that the power spectrum of anisotropic stress of neutrinos evolves as $\eta^{4}$ on superhorizon scales.
After the horizon crossing, the anisotropic stress of neutrinos on small scales
undergoes the free-streaming. This is because the sources of the
anisotropic stress of neutrinos, i.e., $\mathcal{N}^{(1)}_{\ell \geq 1,
0}$, also undergo the free-streaming and the first order gravitational
potentials decay away in the radiation dominated epoch. 
However, on large scales, the anisotropic stress does not decay in the
matter dominated epoch because it can be sustained by the sources
consist of the scalar gravitational potentials that are constant in time
in the matter dominated epoch.
\subsection{The power spectrum of secondary GWs}
\begin{figure}[h]
\begin{center}
\rotatebox{0}{\includegraphics[width=0.7\textwidth]{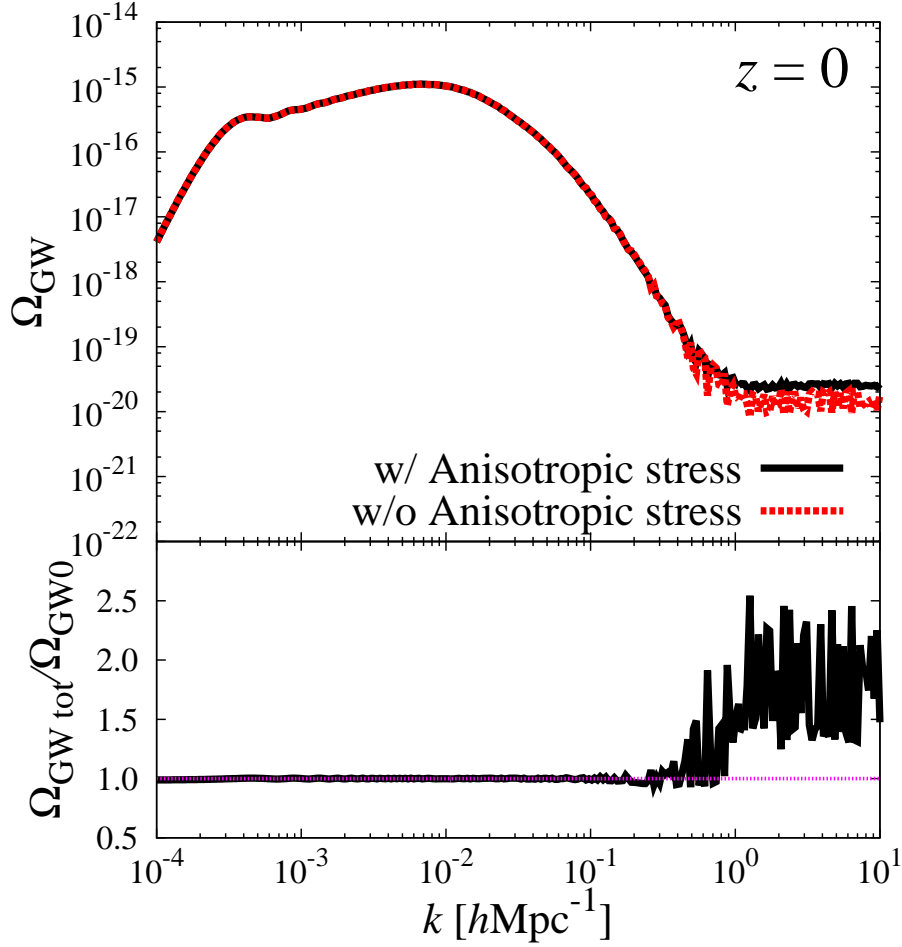}}
\end{center}
\caption{
The secondary GWs at $z = 0$ induced by all the contributions, i.e., the products of the first-order scalar-perturbations and prey second-order anisotropic stresses of photons and neutrinos (black line in the top panel).
The top panel shows the generated GWs with and without the contributions from the purely second-order anisotropic stress at $z=0$ as indicated.
The bottom panel shows the ratio of the spectra with ($\Omega_{\rm GW~tot}$) and without ($\Omega_{{\rm GW} 0}$) the contributions from the purely second-order anisotropic stress.
For reference, the ratio equal to unity is shown (magenta, short dashed line).
The spectrum of GWs for $k\lesssim 1.0~h{\rm Mpc}^{-1}$ at $z = 0$ is consistent with the result of Ref.~\cite{Baumann:2007zm} that was derived without contributions from the purely second-order anisotropic stress.}
\label{fig:total GWs}
\end{figure}

Finally, we show the complete spectrum of the secondary GWs by taking into account all the contributions at second order, namely, the products of the first-order scalar-perturbations and purely second-order anisotropic stresses of photons and neutrinos in Fig.~\ref{fig:total GWs}.
As shown in Fig.~\ref{fig:total GWs}, the anisotropic stresses of photons and neutrinos as a whole amplify the secondary GWs by about $120\%$ on small scales, $k\gtrsim 1.0~h{\rm Mpc}^{-1}$.
However, on large scales, $k \lesssim 1.0~h{\rm Mpc}^{-1}$, anisotropic stresses of photons and neutrinos do not affect the secondary GWs.
In the following section, we discuss the impact of the anisotropic stress on the secondary GWs for photons and neutrinos, separately.
\section{Discussion}
\begin{figure}[h]
\begin{center}
\rotatebox{0}{\includegraphics[width=0.45\textwidth]{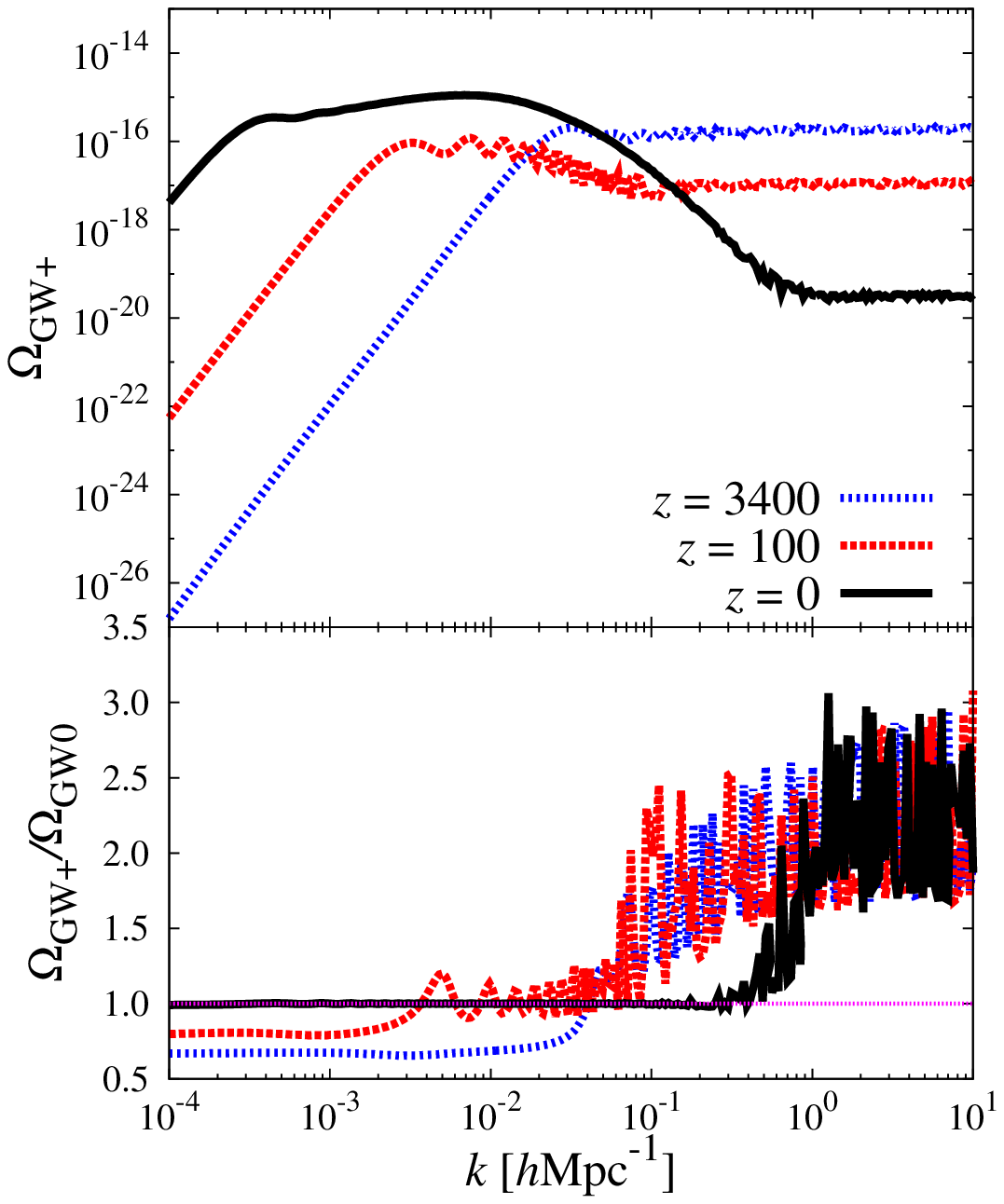}}
\rotatebox{0}{\includegraphics[width=0.45\textwidth]{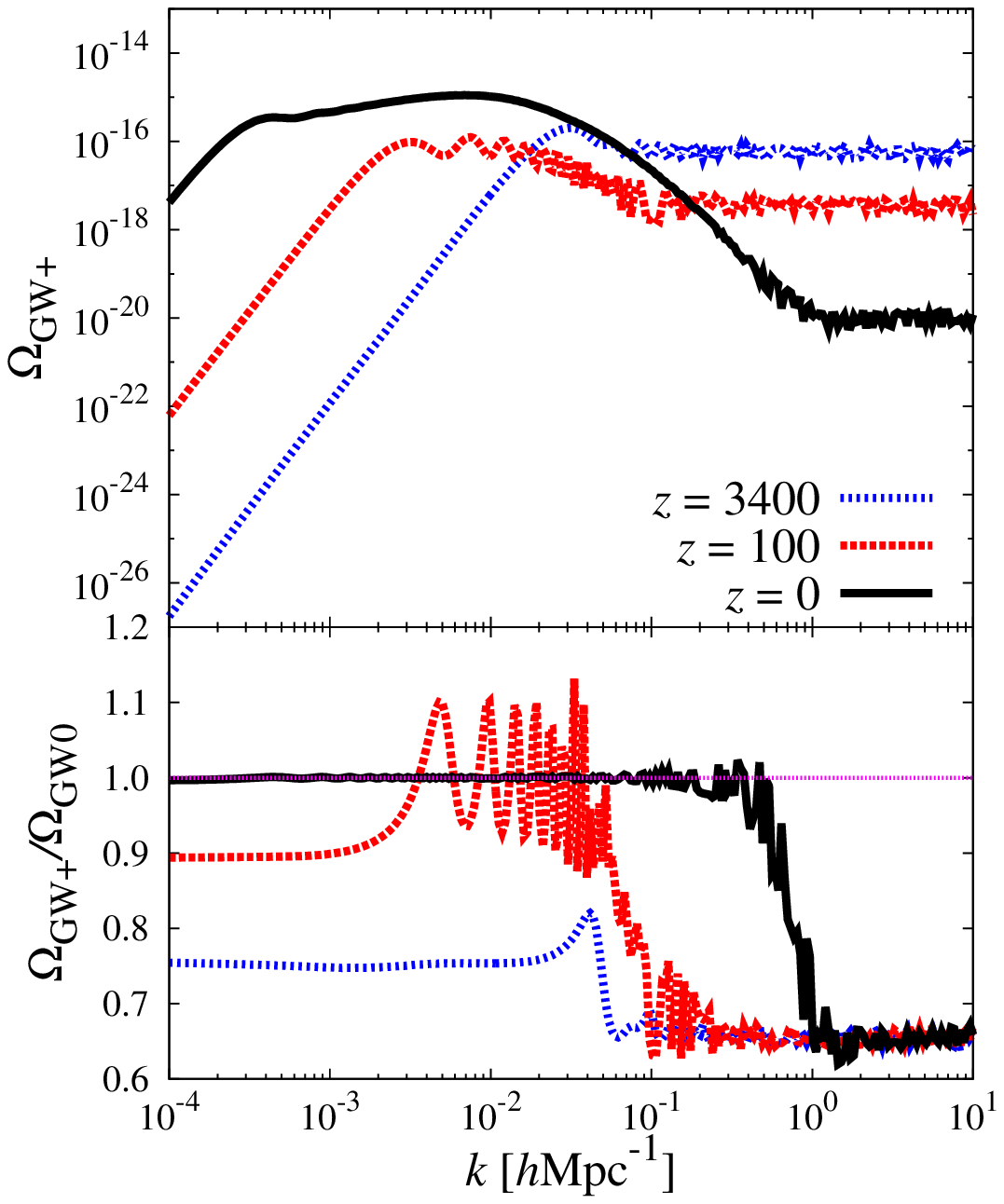}}
\end{center}
\caption{Impact of the anisotropic stresses of photons (\textit{left}) and neutrinos (\textit{right}) on the GWs.
Top figures show the generated GWs including all the contributions at each time as indicated.
Bottom figures show the ratio of with ($\Omega_{\rm GW +}$) and without ($\Omega_{{\rm GW}0}$) the anisotropic stress contributions.
For reference, the ratio equal to unity is shown (magenta, short dashed line).
}
\label{fig:GWs results}
\end{figure}

We discuss effects of second-order anisotropic stress
on the gravitational wave spectrum. We depict the spectrum of
GWs and the ratio of the spectra
with and without the purely second-order anisotropic stress in
Fig.~\ref{fig:GWs results}.

First, we focus on small scales, say, $k \gtrsim 1.0~h{\rm
Mpc}^{-1}$. 
We find that the photon anisotropic stress amplifies GWs about $150\%$ but neutrino anisotropic stress suppresses GWs about $30\%$.
The net effect of purely second-order anisotropic stress on small scales is the
amplification of GWs by about $120\%$ compared to the
case without the photon and neutrino anisotropic stresses.

\begin{figure}[h]
\begin{center}
\rotatebox{0}{\includegraphics[width=0.49\textwidth]{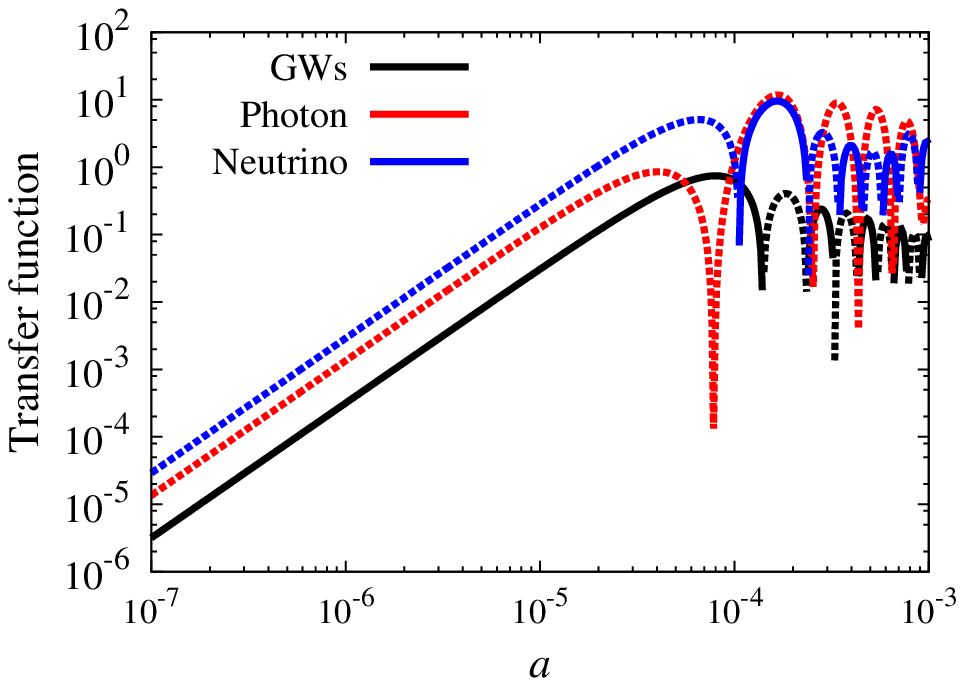}}
\end{center}
\caption{
Transfer functions of anisotropic stresses for photons (red), neutrinos (blue) and GWs without anisotropic stress (black) at $k\approx k_{1}\approx k_{2} \approx 0.1~h{\rm Mpc}^{-1}$.
In this figure, solid lines and dashed lines represent plus and minus signs, respectively.
}
\label{fig: transfer}
\end{figure}

To understand why photon and neutrino give the opposite effects on the GWs, we show the second-order transfer function of anisotropic stress at $k\approx k_{1}\approx k_{2}$, in Fig.~\ref{fig: transfer}, which scales mainly contribute to the second-order power spectrum (see section~\ref{sec: triangle}). 
As is seen in Fig.~\ref{fig: transfer}, before the horizon crossing, the transfer functions for photons and neutrinos have minus sign.
However, after the horizon crossing, the transfer functions for photons and neutrinos have different signs.
Neutrino anisotropic stress undergoes free-streaming after the horizon crossing and starts to oscillate.
In contrast, from Eq.~(\ref{CPT2TCA0 pi g}),
the photon anisotropic stress is sourced by the term proportional to $v^{(1)}_{\gamma~0}(k_{1})v^{(1)}_{\gamma~0}(k_{2})\mathcal{Y}^{1,1}_{2,\sigma}(\hat{k}_{1},\hat{k}_{2})$ due to the scattering term.
For the most contributing configuration where $\bm{k} = \bm{k}_{1} = \bm{k}_{2}$, the first terms, $v^{(1)}_{\gamma~0}v^{(1)}_{\gamma~0}$ do not change the overall sign.
On the other hand, the second term, $\mathcal{Y}^{1,1}_{2,\sigma}(\bm{k}=\bm{k}_{1}=\bm{k}_{2})$ gives a negative sign from Eq.~(\ref{eq: calY}).
The net sign of the photon anisotropic stress is therefore negative through its evolution in the tight-coupling regime.
After the horizon crossing, photons amplify the GWs since the sign of photon anisotropic stress is same as that of the GWs, while neutrinos gives the opposite result.

Second, we focus on GWs on large scales in the matter dominated
epoch. On these scales effects 
from the second-order anisotropic stress should be negligible as the
first order gravitational potentials directly sustain the second-order
GWs.
The transition scales correspond to those above which the
sources from the products of first order perturbations, i.e., the
gravitational potentials in the matter dominated epoch, can sustain the
GWs. On those scales, the gravitational potentials at first order
completely determine the amplitude of GWs and hence the contributions
from purely second-order anisotropic stress become negligible.
More specifically, from Eq.~(\ref{eq: GWs}), the GWs are sourced by $8\pi G\rho^{(0)}_{\gamma} \Delta^{(2)}_{2,\sigma}$ and $8\pi G\rho^{(0)}_{\nu} \mathcal{N}^{(2)}_{2,\sigma}$. In the matter dominated epoch, the prefactor of anisotropic stress, namely $\rho^{(0)}_{\gamma}$ and $\rho^{(0)}_{\nu}$, must be negligibly small.
Thus, in the matter dominated epoch, the anisotropic stress does not affect the GWs.
In the radiation dominated epoch, on the other hand, the anisotropic stress affects
the second-order GWs.
From Fig.~\ref{fig: transfer}, the photon and neutrino anisotropic stresses contribute to the GWs as the negative source.
Consequently, in the radiation dominated epoch, the amplitude of the GWs is suppressed on large scales.
However, these suppressions do not affect final results since the main contributions are determined around the horizon crossing.

\begin{figure}[h]
\begin{center}
\rotatebox{0}{\includegraphics[width=0.45\textwidth]{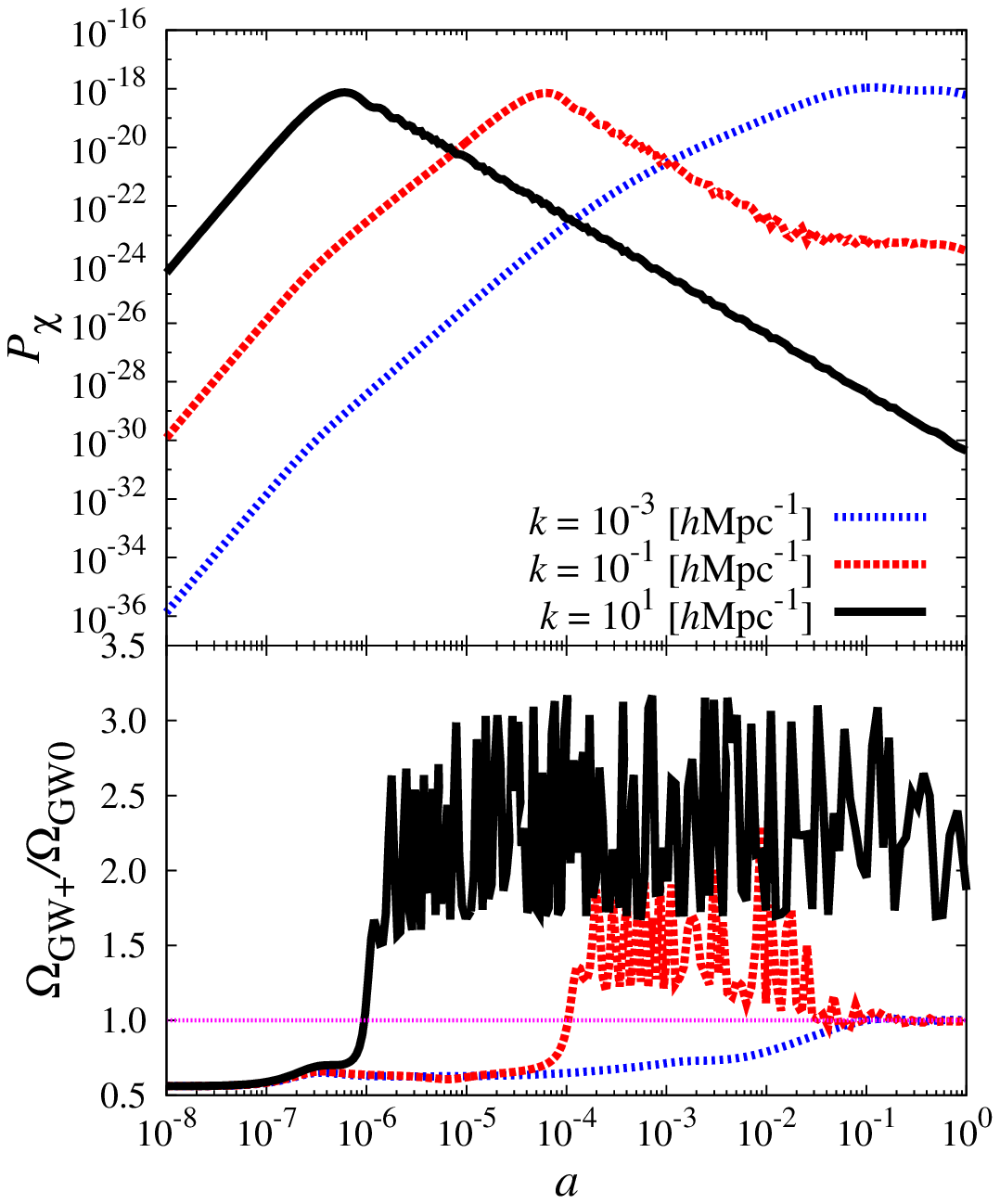}}
\rotatebox{0}{\includegraphics[width=0.45\textwidth]{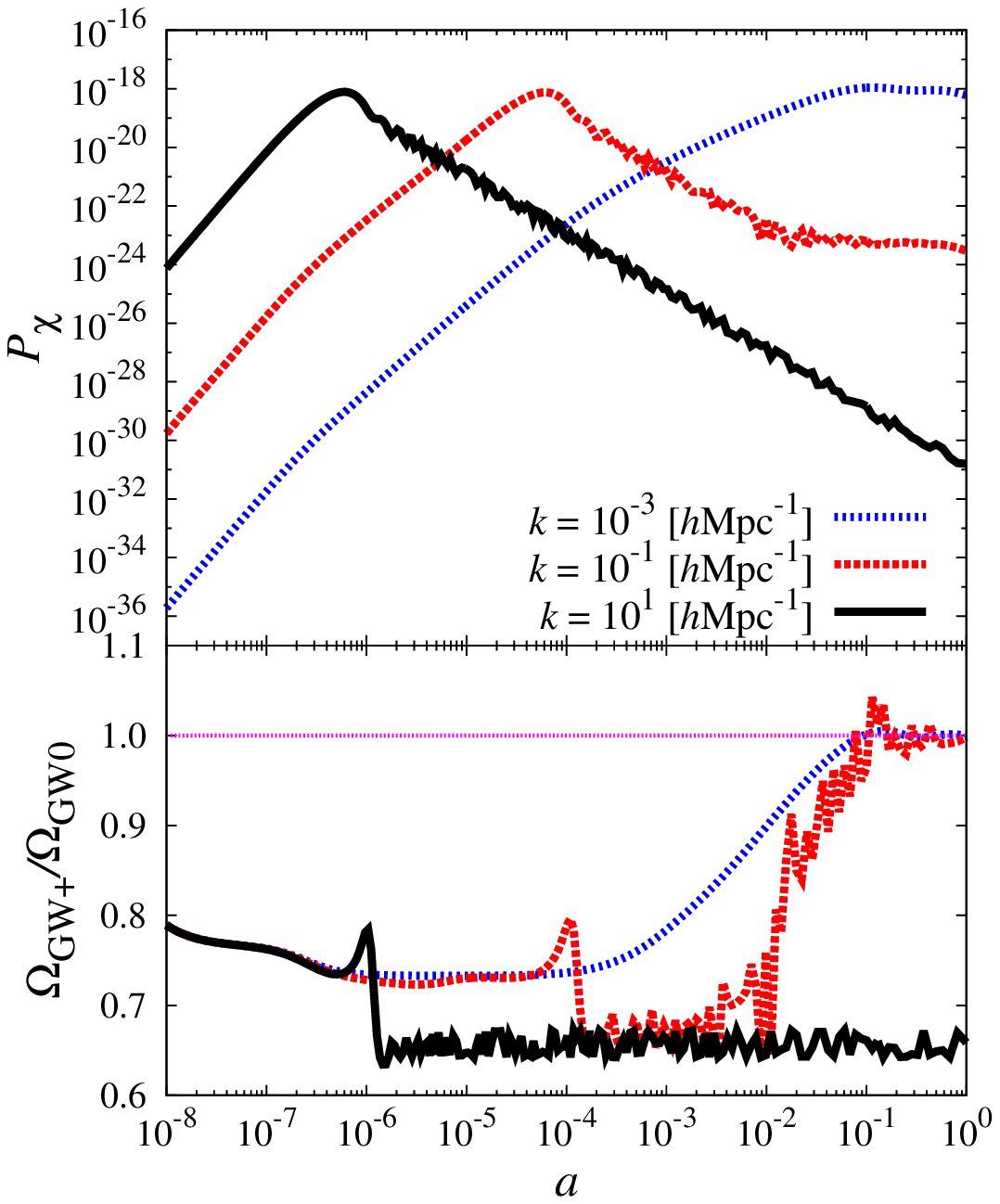}}
\end{center}
\caption{
The effects of the anisotropic stresses of photons (\textit{left}) and neutrinos (\textit{right}) on the evolutions of the GWs.
Top figures show evolutions of generated GWs including all contributions at each scale, $k = 10^{-3}~h{\rm Mpc}^{-1}$, $10^{-1}~h{\rm Mpc}^{-1}$, and $10^{1}~h{\rm Mpc}^{-1}$.
Bottom figures show the ratio of the spectra with ($\Omega_{\rm GW +}$) and without ($\Omega_{{\rm GW}0}$) the anisotropic stress.
For reference, the ratio equal to unity is shown (magenta, short dashed line).
These evolutions have been discussed in Ref.~\cite{Baumann:2007zm} without the second-order anisotropic stress.
}
\label{fig: diff evolve}
\end{figure}

To confirm the above analytic investigations, we show the difference between time evolutions of the GWs with and without anisotropic stress in Fig.~\ref{fig: diff evolve}, for three typical scales, $k = 10^{-3}~h{\rm Mpc}^{-1}$, $10^{-1}~h{\rm Mpc}^{-1}$, and $10^{1}~h{\rm Mpc}^{-1}$.
Before the horizon crossing, the GWs are generated by the product of
first-order scalar perturbations and the purely second-order anisotropic stress.
After the horizon crossing, the power spectrum of GWs decay as $a^{-2}$ because the product of
the first-order scalar perturbations decay faster after the horizon crossing.
However on large scales, i.e., $k\lesssim 10^{-1} ~h{\rm Mpc}^{-1}$, the product of
the first-order scalar perturbations remains constant and sustains the second-order GWs.
Consequently, for large scales where the product of the first-order scalar perturbations dominates to generate the second-order GWs, the effect of anisotropic stress of free-streaming particles is negligible.
These evolutions are consistent with the spectra of GWs in Fig.~\ref{fig:GWs results}.

The reason why we only consider up to the second order perturbation is as follows. 
First of all, GWs from density perturbations can arise only from the second order.
Therefore it is essential to consider the second order perturbations.
Furthermore, from Fig.~\ref{fig: diff
evolve}, we can see that the anisotropic stress
affects the GWs only at the horizon crossing where the perturbation
approach is valid even on smaller scales if the power spectrum of the
primordial density perturbations is nearly scale invariant. After the horizon crossing, 
the second order anisotropic stress should decay away together with its
sources, i.e.,  the first order gravitational potentials in the
radiation dominated era.  Therefore, the
higher order contributions, e.g., those from the third order,
should be small compared with those from the second order on small scales.

Finally, let us discuss observational implications of our result. Cosmological
tensor perturbations are known to induce curl modes of weak gravitational
lensing of background galaxies \cite{2003PhRvL..91b1301D}. It is found that the secondary tensor mode
generated from density perturbations produces larger signal in the curl modes
than the gravitational waves generated by inflation if the tensor-to-scalar
ratio is less than $0.4$ \cite{2008PhRvD..77j3515S}. The results presented in
that article should remain unchanged because the purely second order
anisotropic stress considered here does not alter the amplitude of GWs on
large scales relevant to the weak gravitational lensing signal.

On small scales, the amplitude of the second-order GWs can be
larger than that of the GWs from inflation 
if the
tensor-to-scalar ratio is $r\lesssim 10^{-4}$, and becomes relevant in
direct detection 
experiments such as DECIGO~\cite{Seto:2001qf} and atomic gravitational
wave interferometric sensors~\cite{PhysRevD.78.122002}.  
In this case, the second-order GWs may be directly observed, and they can be used as yet another probe of standard ΛCDM cosmology.
For example, it is claimed that DECIGO may detect the stochastic GWs background up to the order of
$\Omega_{\rm GW}\sim 10^{-20}$ at $f\sim 0.1~{\rm Hz}$ (\cite{Seto:2001qf}).  In our numerical
calculations, we find that the second-order GWs have the amplitude
about $\Omega^{({\rm 2nd})}_{\rm GW}\sim f_{\Pi}\times 10^{-20}$ on
small scales, with $f_{\Pi}$ being the amplification factor by the
second-order anisotropic stress.  Therefore, the estimated amplitude of the
secondary GWs is at the same order of magnitude as the noise level assumed for DECIGO~\cite{Seto:2001qf}, and it will become important for the
future GWs experiments to estimate the impact of the anisotropic
stress on the secondary GWs.

Furthermore, the secondary induced tensor mode has been used to place
constraints on the amplitude of primordial scalar perturbations, which can not
be probed by other ways
\cite{2009PhRvL.102p1101S,2010PhRvD..81b3527A,2010PThPh.123..867S}. The
reason is that the amplitude of the secondary induced tensor mode is
proportional to $P_{\rm scalar}^2$, and therefore non-detection of GWs in
gravitational wave experiments places an upper bound on $P_{\rm scalar}$ at
observed scales. If we assume that our result applies to much smaller scales
$k\gg 100$ Mpc$^{-1}$, the constraints on $P_{\rm scalar}$ should become
tighter by a factor of $1.5$, because our result shows that on small scales
the secondary induced tensor mode should be larger by a factor of $2.2$ than the
previous estimates. In fact, in the very early universe with temperature
above $\sim 1$ MeV, neutrinos can not stream freely and we expect that
anisotropic stress of neutrinos also contributes to the enhancement of the
secondary induced gravitational waves in the same way as photons. 
In the tight coupling limit, anisotropic stresses of neutrinos and
photons should give 
similar effects on the secondary GWs, because the amplitude of
anisotropic stress at second order in the tight 
coupling limit does not depend on the tight coupling parameter
explicitly, as shown in Eq.(25). In this case, the amplitude of secondary induced GWs
should be larger by $150 \% /f_\gamma \sim 252 \%$ compared with the
case neglecting the anisotropic stress, where $f_\gamma \approx 0.681$
is the energy fraction of photons in the radiation dominated era with
temperature $\lesssim $ MeV. Other relativistic particles
such as electrons and positrons are 
also expected to amplify the GWs even further for $k\gg 100$ Mpc$^{-1}$.
\section{Summary}
In this paper, we explored the impact of anisotropic stresses of
photons and neutrinos on the secondary GWs generated from first order density perturbations. To estimate
the spectrum of GWs including the anisotropic stress, we
reformulated the cosmological perturbation theory up to second-order,
which is based on the Einstein and Boltzmann equations. To solve the
second-order 
equations numerically, we considered the initial conditions of
the Boltzmann equation of photons using the tight coupling approximation.
Deep in the radiation dominated era, photons and baryons frequently
interact with each other, which allows us to expand the equations using the tight-coupling
parameter $k/\dot{\tau}_{c}$. Under the tight-coupling approximation,
we solved the hierarchical Boltzmann equations and showed the
tight-coupling 
solution up to first-order. In the first-order cosmological
perturbation theory, the anisotropic stress of photons must vanish due 
to frequent
collisions between photons and baryons, as well as the photon's higher
multipoles. However in the second-order cosmological perturbation theory, the
anisotropic stress of photons does not vanish because the mode coupling of
the first-order scalar perturbations, i.e., velocity perturbations of
photons, generates anisotropic stress.
Moreover, differently from the first order perturbation theory,
the octupole moment ($\ell = 3$) of the distribution function of photons
can arise even in the first-order tight-coupling solution. We found that
this is because the octupole moment is also sourced 
by the mode coupling of the first-order scalar perturbations and the streaming
term.
In the next order of the tight-coupling expansion, we 
showed that the relative velocity between photons and baryons does not
vanish, which is the same result as in the first-order perturbation
theory. We adopted these solutions as the initial conditions for second order perturbations, and solved
the Einstein-Boltzmann system numerically.

Photons and massless neutrinos have anisotropic
stress in the perturbed universe. We considered effects of
anisotropic stresses of photons and neutrinos on the secondary generated
GWs from density perturbations. 
We found that, before the 
recombination epoch, anisotropic stresses of photons and neutrinos
affects the amplitude of GWs on both small and large scales. On
super horizon scales, the anisotropic stress is growing because of the 
first-order scalar perturbations. Generated anisotropic stress can be
a source of the GWs. 
In the matter
dominated epoch, the effect of anisotropic stress on the GWs becomes negligible,
because the first-order scalar gravitational potentials directly sustain the GWs. 
On small scales, on the other hand, the GWs have been affected by the
anisotropic stress because the scalar
gravitational potentials have decayed away on small scales. 
Photon anisotropic stress, which is sourced by the square of the first-order velocity perturbations, amplifies the GWs by about $150\%$.
On the other hand, neutrino anisotropic stress suppresses the GWs by about $30 \%$ because the second-order neutrinos undergo the free-streaming with oscillations around the origin.
This difference can be explained by the different signs of the transfer functions.
To conclude,
the effect of the anisotropic stress at second-order in cosmological
perturbations is to amplify the GWs by about $120$ \% for $k\gtrsim
1.0~h{\rm Mpc}^{-1}$ in the present universe, compared to
the case without taking the anisotropic stress into account. 

\begin{acknowledgments}
This work was supported in part by a Grant-in-Aid for JSPS Research under Grants No.~26-63 (SS), and a Grant-in-Aid for JSPS Grant-in-Aid for Scientific Research under Grants No.~24340048 (KI) and 25287057 (NS).
We also acknowledge the Kobayashi-Maskawa Institute for the Origin of Particles and the Universe, Nagoya University, for providing computing resources useful in conducting the research reported in this paper.
\end{acknowledgments}
\bibliography{ref}
\end{document}